\documentclass[preprint,superscriptaddress,
 amsmath,amssymb,
 aps,
fleqn,nofootinbib, groupedaddress
]{revtex4-1}

\usepackage{ucs}
\usepackage[utf8x]{inputenc}
\usepackage{ae}
\usepackage{psfrag}
\usepackage{dcolumn}
\usepackage{bm}
\usepackage{mathrsfs}
\usepackage{graphicx}
\usepackage{color}
\usepackage{pstricks}
\usepackage{xcolor,rotating,epic,eepic}
\usepackage{multirow}
\usepackage{dcolumn}
\usepackage[caption=false]{subfig}
\newcommand{\gr}{\boldsymbol}
\newcommand{\ds}{\displaystyle}

\begin{document}


\author{Elie Favier}

\author{Navid Nemati}
\email{navid.nemati@u-pem.fr}
\email{nnemati@mit.edu} 

\author{Camille Perrot}

\author{Qi-Chang He}
\email{qi-chang.he@u-pem.fr}
\address{Laboratoire de Mod\'elisation et Simulation Multi Echelle, Universit\'e Paris-Est, UMR 8208 CNRS, 77454 Marne-la-Vall\'ee, France}


\title{Generalized analytic model for rotational and anisotropic metasolids}

\begin{abstract}

An analytical approach is presented to model a metasolid accounting for anisotropic effects and rotational mode. The metasolid is made of either cylindrical or spherical hard inclusions embedded in a stiff matrix via soft claddings
. It is shown that such a metasolid exhibits negative mass densities near the translational-mode resonances, and negative density of moment of inertia near the rotational resonances. As such, the effective density of moment of inertia is introduced to characterize the homogenized material with respect to its rotational mode. The results obtained by this analytical and continuum approach are compared with those from discrete mass-spring model, and the validity of the later is discussed.  Based on derived analytical expressions, we study how different resonance frequencies associated with different modes vary and are placed with respect to each other, in function of the mechanical properties of the coating layer. We demonstrate that the resonances associated with additional modes taken into account, that is, axial translation for cylinders, and rotations for both cylindrical and spherical systems, can occur at lower frequencies compared to the previously studied plane-translational modes.



\end{abstract}



\maketitle


\section{Introduction}
\label{section introduction}

Elastic or acoustic metamaterials  are structures with subwavelength units that exhibit unusual  macroscopic parameters within certain frequencies, such as negative mass density \cite{sheng2000,sheng2005, wu2008}, negative elastic bulk modulus \cite{fang2006,nemati2015},  simultaneous  negative mass density and elastic bulk modulus \cite{chan2004, ding2007, cheng2008, lee2010, liu2011, sheng2013}, or negative density and shear modulus \cite{wu2011,sheng2011}, and negative index of refraction \cite{brunet2015}. Metamaterials with unconventional constitutive parameters have broadly extended the ability of manipulation and control of mechanical wave propagation.  Wave control can arise from the formation of band gaps, that are the frequency bands where the propagation becomes forbidden. Differently from the Bragg band-gaps in phononic crystals \cite{sigalas1993, djafari1993} that are produced based on the collective effects of the periodically-arranged scatterers in the medium, the spectral gaps manifested in metamaterials originates in localized resonances of the material building-blocks. Thus, in metamaterilals the production of  stop bands depends on the internal structure of the building units. Another mechanism emerged to achieve extreme material-parameters is based on space coiling \cite{liang2012, cummer2013}.

Over the last decade, research on anisotropic metamaterials has enhanced the ability to control sound and elastic wave propagation, leading to the proposal and fabrication of new devices \cite{milton2007, cummer2007, chan2008, zhang2009, christensen2010, zhu2011, yu2014, djafari2015}.  The first acoustic metamaterial that was realized is a block of stiff material (epoxy) with uniformly (or periodically) dispersed  locally resonant  structural units that consist of a high-density lead sphere coated with a soft material. This  material exhibits resonance-based spectral-gaps in low-frequency regime where the sonic wavelength is of two orders of magnitude larger than the lattice constants. Thus, this designed material was a major progress as sound attenuation is challenging in low frequencies. An analytical model has been proposed to capture the essence of physics dealing with this type of materials \cite{sheng2005}. Since in practice the lead sphere and the host are much stiffer compared with the coating material, in this model the host and coated sphere have been approximated to be rigid within the long-wavelength limit.  Considering the two-dimensional (2D) system with coated cylindrical inclusions, and three-dimensional (3D) with coated spheres, effective dynamics of the metasolid was derived in resonance-induced band-gap regimes related to the translational motions.  Analytical expressions were given for the effective mass density which becomes negative near the resonance.

In this paper, generalizing the simple model proposed in Ref. \cite{sheng2005}, we describe the dynamics of the 3D three-component composite taking into account the rotational motions of the lead cylinder or sphere, and the matrix. It is shown that when the inclusions are cylindrical, because of their geometrical anisotropy, wave propagation in the direction perpendicular to the axis of cylinders gives rise to different effective density in comparison  with the effective density describing the propagation perpendicular to that axis.  Components of the anisotropic effective density are given through analytical expressions. Furthermore, we introduce and analytically calculate the effective density of the moment of inertia, as a parameter that describes the effective rotational mode of the material. It has been previously demonstrated that, when the cladding is very soft, the stop-band frequency can be very low \cite{sheng2000}. This makes such a composite with a practical size interesting for various applications including sound and vibration isolations. Here, allowing for additional degrees of freedom (DOFs) related to rotational motions and translation along cylinder axis, we show that the model predicts spectral gaps characteristic for each mode of propagation. In particular, with the material components chosen in Ref. \cite{sheng2005}, we demonstrate that in the case of spherical inclusions, the band gaps related to rotational mode occur in lower frequencies compared with those corresponding to translational modes, where the respective effective parameters become negative. Also, for the case of cylindrical inclusions,  comparison between the stop bands for translational modes along  and perpendicular to the cylinders shows that  those that occur along the cylinder axis are formed at lower frequencies. In general, for materials made of spherical or cylindrical inclusions, we analyze the occurrence of local resonances related to each mode on the frequency axis, as function of coating-material properties. It turns out that accounting for additional DOFs that results in additional propagation modes improve notably the tunability of the material in terms of size and relevant frequencies. This is significant for applications such as sound and vibration proofing, which requires small-size materials enabling the low-frequency wave attenuation.

Rotational modes in periodic solid composites have been previously studied \cite{milton2007, zhao2005, liu2011, peng2012, guenneau2013, huang2014, peng2013}.  Rotary resonances have been estimated by a simple model in 2D  square lattice of glass cylinders in epoxy \cite{zhao2005}.  Numerical analysis was performed to obtain double negative properties, for density and bulk modulus,  by utilizing simultaneously the local translational and rotational resonances of a chiral microstructure with a unit cell of three-component solid media composed of a chirally soft-coated heavy cylinder core embedded in a elastic matrix \cite{liu2011}. Periodic system of rotating resonators coated by an  anisotropic heterogeneous elastic material embedded in a matrix, which was modeled by an asymptotic approach, has been reported to produce negative refraction \cite{guenneau2013}. Thus, by modelling the anisotropic material that is placed between the core and matrix as ligaments, and by coupling the translational motions with rotations, it has been numerically demonstrated  that this structured medium could be employed for lensing and mode localization. Due to simultaneous translational and rotational local-resonances,  a metasolid  made of a chiral microstructure with a single phase material served as a system to experimentally achieve negative refraction \cite{huang2014}.  Accounting for rotational modes, a mass-spring model was used to reproduce band gaps in a 2D phononic crystal made of rigid cylinders in epoxy \cite{peng2012}. This model was enhanced to describe dispersion relation for rotational modes of a 2D square array of rubber-coated steel cylinders in epoxy \cite{peng2013}. However,  no analytical approach for rotational metasolids based on full elastodynamics 
 has been reported so far.

Here, we study rotational modes and analyze their associated homogenized properties by modeling analytically the classical three-component subwavelength structures through taking general assumptions and using direct first-principle elastodynamics without any fitting parameters. This offers us not only essential physical insights for the effective dynamics of the metasolid but also a quick and efficient guide to design materials with local rotational-resonances. Additionally and in parallel, for each mode we have systematically calculated the effective parameters of the metasolid based on the discrete mass-spring modeling, and compared the results with those arising from the continuum elastodynamic model. This comparison clarifies the limits of the mass-spring based model, in particular in terms of describing the effective-material dynamics related to the second local-resonance phenomenon that manifests itself for all translational and rotational modes.   

In the following, we introduce the elastodynamics equations at microscale in Sec. \ref{section microscale} for an arbitrary shape of the inclusions. We then study the case of cylindrical inclusions in Sec. \ref{section cylinder}, followed by the analysis for spherical inclusions in Sec. \ref{section sphere}. In Sec. \ref{section homog}, the effective parameters for translational and rotational modes are calculated for the homogenized media including microscopically either cylindrical or spherical resonators. In the concluding Sec. \ref{section conclusion}, the main results of the paper are briefly summarized.


\section{Microscale equations}\label{section microscale}

\begin{figure}
\includegraphics[height=5 cm,draft=false]{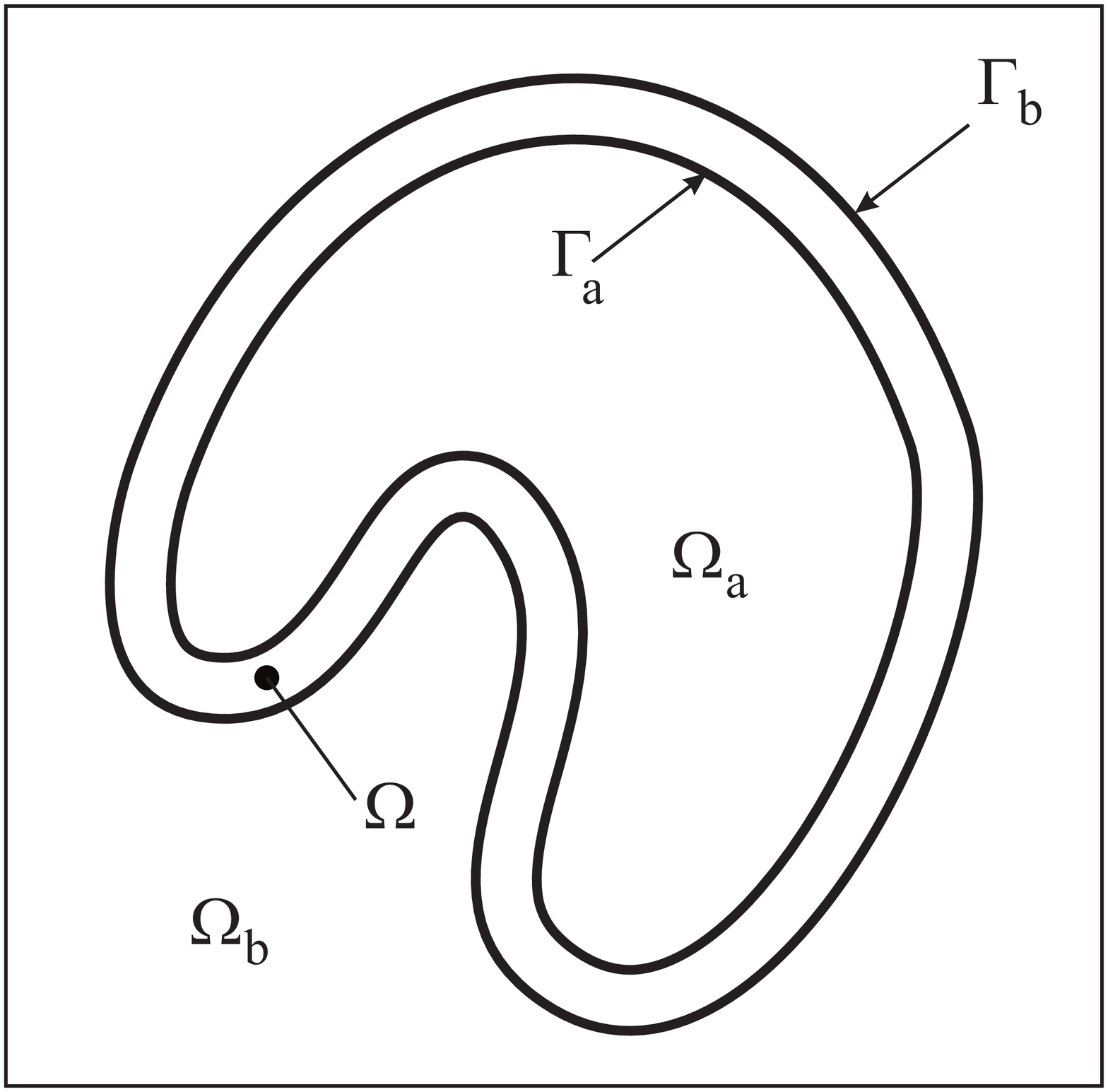} 
\caption{Illustration of the cross-section of a 3D unit cell with coated inclusion of arbitrary shape: rigid inclusion $\Omega_a$ surrounded by the elastic cladding $\Omega$ and embedded in the rigid matrix $\Omega_b$. The boundary regions of the cladding are denoted by $\Gamma_a$ and $\Gamma_b$.}
\label{geometry}
\end{figure}

The unit cell under consideration consists  of a hard inclusion $\Omega _{a}$ embedded in a hard matrix $\Omega_{b} $ via a soft cladding $\Omega $ (Fig. \ref{geometry}). The interfaces $\Gamma _{a}$ and $\Gamma_{b}$ located between $\Omega _{a}$ and $\Omega $, and  between $\Omega_{b}$ and $\Omega $, respectively, are assumed to be perfect. The materials constituting $\Omega _{b}$ and $\Omega _{a}$ are very stiff with respect to the material forming $\Omega $. For this reason, and within the long-wavelength limit,  we make the assumption that the inclusion $\Omega _{a}$ and matrix $\Omega _{b}$ are rigid while the cladding $\Omega $ is linearly elastic \cite{sheng2005, bonnet2015}.

The elastodynamic formulation of the unit-cell system is based on the assumptions that only the cladding is deformable and that  the displacements of the inclusion  and matrix  are small and the strains inside the soft layer are infinitesimal. Consequently, the kinematics of the inclusion and the matrix can be formulated by
\begin{equation}
\bm{\hat{u}}_{\alpha}(\bm{r},t)=\bm{\hat{V}}_{\alpha}(t)+\bm{\hat{\Theta}}_{\alpha}(t)\times \bm{r},\; \text{on}\, \, \Gamma_\alpha,  \label{RM}
\end{equation}
where $\alpha=a,b$, $\bm{r}$\ is the position vector relative to a the origin $O$ of a 3D space, and $\bm{\hat{u}}$ represents the displacement field generated by a translational displacement vector $\bm{\hat{V}}_{\alpha}$ and an infinitesimal rotation characterized by the 3D vector $\bm{\hat{\Theta}}_{\alpha}$. Considering the time-harmonic motion of the foregoing unit system, we write $\bm{\hat{u}}(\bm{r},t)=  \bm{u}(\bm{r},\omega )e^{ i\omega t}$, $\bm{\hat{V}}_{\alpha}(t)=\bm{V}_{\alpha}(\omega)e^{ i\omega t}$, and $\bm{\hat{\Theta}}_{\alpha}(t)= \bm{\Theta }_{\alpha}(\omega )e^{ i\omega t}$, where $i=\sqrt{-1}$ and $\omega $  is the angular frequency. The motion of the cladding  made of a linearly elastic isotropic homogeneous material is described by the following Navier equation together with the displacement continuity across the perfect interfaces $\Gamma _{b}$\ and $\Gamma _{a}$: 
\begin{subequations}\label{clading governing}
\begin{eqnarray}
\hspace*{-7mm}\lefteqn{(\lambda +2\mu )\bm{\nabla }({\bm{\nabla }}.\bm{u})-\mu \bm{\nabla }\times (\bm{\nabla }\times \bm{u})=-\rho \omega ^{2}\bm{u},\, \, \text{in}\,  \Omega} \label{Navier-w}\\
\hspace*{-7mm}\lefteqn{\bm{u}(\bm{r},\omega )=\bm{V}_{\alpha}(\omega )+\bm{\Theta }_{\alpha}(\omega )\times \bm{r},\, \,  \text{on}\, \, \Gamma_\alpha,\, \alpha=a,b}  
\label{BC-w}
\end{eqnarray}
\end{subequations}
where $\lambda $ and $\mu $ are the Lam\'{e} constants and $\rho $ is the mass density of the coating material.

For later use, it is convenient to introduce the Helmholtz decomposition:
\begin{equation}
\bm{u}=\bm{\nabla }\Phi +\bm{\nabla }\times  \bm{\Psi}, \hspace*{5mm} \bm{\nabla }.\bm{\Psi}=0,  \label{Hequ}
\end{equation}
where $\Phi $ is a scalar field and $\bm{\Psi}$ a vector field. Using Eq. (\ref{Hequ}),  the equation (\ref{Navier-w}) can be recast into the following ones
\begin{subequations}\label{helmholtz equations}
\begin{eqnarray}
\lefteqn{\bm{\nabla}^{2} \Phi +h^{2}\Phi =0,}  \label{H1}\\
\lefteqn{\bm{\nabla}^{2}\bm{\Psi}+\kappa ^{2}\bm{\Psi}=\bm{0}}
\label{H2}
\end{eqnarray}
\end{subequations}
where $h=\omega/c_{l}$ and $\kappa =\omega/c_{t}$ with  $c_{l}=\sqrt{(\lambda +2\mu)/\rho }$  and $c_{t}=\sqrt{\mu /\rho }$ being  the celerities of the longitudinal and transverse elastic waves, respectively, in an infinite linearly elastic isotropic medium. In the following, we shall analytically solve the Helmholtz equations (\ref{helmholtz equations}) with the boundary conditions (\ref{BC-w}) for two special cases of major interest.

%

\section{Media with cylindrical inclusions}\label{section cylinder}
  
\begin{figure}
\includegraphics[height=4cm,draft=false]{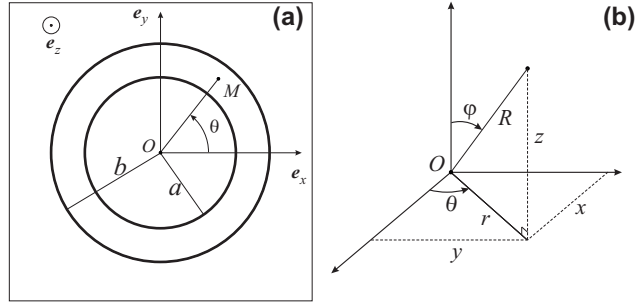} %
\caption{Illustration of a unit cell with cylindrical or spherical inclusion in $xy$ plane (a), and coordinate system in 3D space (b).}
\label{fig:coord}
\end{figure}

The first special case concerns a material with structural units consisting of three coaxial cylinders of height $L$, uniformly or periodically distributed. The inner cylinder $\Omega _{a}$ is a long revolution one of radius $a\ll L$ while the outer cylinder $\Omega _{b}$ is bounded by an inner cylinder of circular cross-section of radius $b>a$ and an outer cylindrical surface with square cross-section (Fig. \ref{fig:coord}a). 

\begin{figure}
\includegraphics[height=4.3cm,draft=false]{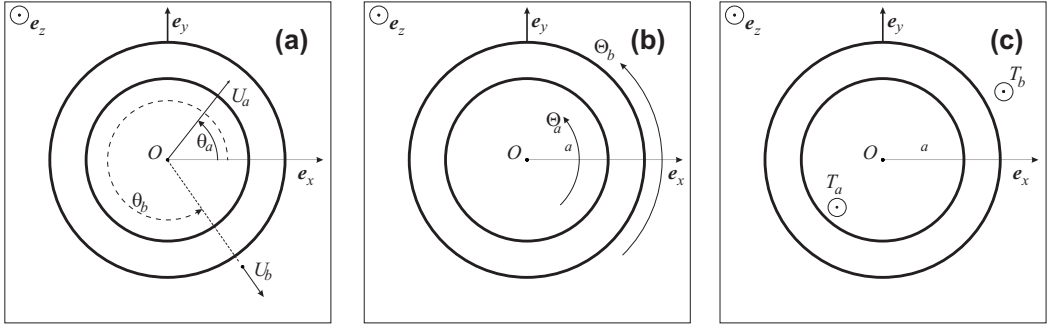}
\caption{Illustration of boundary conditions at the interfaces of the cladding, $r=a$ and $r=b$, for plane translation (a), rotation (b),  and axial translation (c) motions.}
\label{fig:CL}
\end{figure}

Regarding the geometry of the unit cell, it is convenient to use the cylindrical coordinates associated orthonormal vectors $\bm{e}_{r}$, $\bm{e}_{\theta }$, and $\bm{e}_{z}$  (Fig. \ref{fig:coord}b). As it was assumed in Sec. \ref{section microscale}, the inner and outer cylinders are rigid, whereas the cylindrical cladding is deformable. In addition, since the dimensions of the unit cell are such that $L\gg b>a$, the six  DOFs of the motion of $\Omega _{a}$ relative to $\Omega _{b}$ can be reduced to four: two translations in the transverse plane, one translation along the cylinder axis and one rotation about the latter. Thus, with no loss of generality, the boundary conditions (\ref{BC-w}) for the cladding $\Omega $ (Fig. \ref{fig:CL}) can be written as ($\alpha=a,b$)
\begin{eqnarray}
\hspace*{-7mm}\bm{u}|_{r=\alpha} &=&U_{\alpha}\left\lbrace  \cos \theta _{\alpha}\bm{e}_{x}+\sin\theta _{\alpha}\bm{e}_{y}\right\rbrace  
+\alpha\Theta _{\alpha}\bm{e}_{z}\times  \bm{e}_{r}+T_{\alpha}\bm{e}_{z},  \label{CylBC} 
\end{eqnarray}
where $U_{\alpha}$ represents the amplitude of the displacement in the plane direction $\bm{e}_{\alpha}$ resulting from the rotation of $\bm{e}_{x}$  by the angle $\theta _{\alpha}$  around $ \bm{e}_{z}$. The last term in 
 the above expressions stands for the axial translation, with  $T_{\alpha}$  the amplitude of the displacement along $\bm{e}_{z}$. The last DOF, i.e. the rotation around the common axis $\bm{e}_{z}$,  is proportional to $\Theta _{\alpha}\bm{e}_{z}\times \bm{e}_{r}$, where $\Theta_{\alpha}$ is the infinitesimal amplitude of the rotation. To link the general form of the boundary conditions (\ref{BC-w}) to Eq. (\ref{CylBC}) associated with the case of  cylindrical inclusions, it is obvious that $\bm{V}_{\alpha}\equiv U_{\alpha}\bm{e}_{\alpha}+T_{\alpha}\bm{e}_{z}$ and $\bm{\Theta }_{\alpha}\equiv\Theta_{\alpha}\bm{e}_{z}$ (Fig. \ref{fig:CL}).

Considering the cylindrical nature of both the geometry and boundary conditions, the displacement field $\bm{u}$ is clearly independent of the axial coordinate $z$, thereby the potentials $\Phi$ and $\bm{\Psi}$ are also independent of $z$.  Moreover, following Love \cite{Love44}, $\bm{\Psi}$ can be chosen to take the form $\bm{\Psi}=\psi_{\theta}(r,\theta)\bm{e_{\theta}}+\Psi_{z}(r,\theta)\bm{e}_{z}$. The condition $\bm{\nabla }.\bm{\Psi}=0$ in Eq.  (\ref{Hequ}) is now satisfied if and only if $\Psi_{\theta}$ is independent of $\theta$. Thus, the potential $\bm{\Psi}$ can be expressed as $\bm{\Psi}=\Psi_{\theta}(r)\bm{e}_{\theta}+\Psi_{z}(r,\theta)\bm{e}_{z}$. 

The general form of the  solutions to the Helmholtz equations (\ref{helmholtz equations}) can  be written as 
\begin{subequations}\label{solutions general form}
\begin{eqnarray}
\hspace*{-7mm}\lefteqn{\displaystyle\Phi(r,\theta)=\sum_{n=0}^{+\infty}\left\lbrace A_n\mathrm{J}_n(hr)+B_n\mathrm{Y}_n(hr) \right\rbrace \times 
\left\lbrace C_n\cos(n\theta)+D_n\sin(n\theta)\right\rbrace ,} \\
\label{eq:SolSeries_1}
\hspace*{-7mm}\lefteqn{\Psi_z(r,\theta)=\sum_{n=0}^{+\infty}\left\lbrace E_n\mathrm{J}_n(\kappa r)+F_n\mathrm{Y}_n(\kappa r)\right\rbrace \times
\left\lbrace G_n\cos(n\theta)+H_n\sin(n\theta)\right\rbrace ,} \\ 
\label{eq:SolSeries_2}
\hspace*{-7mm}\lefteqn{\Psi_{\theta}(r)=A\mathrm{J}_1(\kappa r)+B\mathrm{Y}_1(\kappa r),}
\label{eq:SolSeries_3}
\end{eqnarray}
\end{subequations}
where $\mathrm{J}_n$ (\textit{resp.} $\mathrm{Y}_n$) stands for the $n^{th}$-order Bessel function of the first (\textit{resp.} second) kind. In these solutions, $A_n$, $B_n$, $C_n$, $D_n $, $E_n$, $F_n$, $G_n$, $H_n$, $A$, and $B$ are unknown constants to be determined. 

Boundary conditions (\ref{CylBC}) and  Helmholtz decomposition (\ref{Hequ}) expressed in cylindrical coordinates lead to
\begin{subequations}\label{BC cylinder}
\begin{eqnarray}
\hspace*{-7mm}\lefteqn{\left[\frac{\partial \Phi}{\partial r}+\frac{1}{r}\frac{\partial \Psi_z}{\partial\theta}\right]_{r=\alpha,\theta}=U_{\alpha}\cos( \theta _{\alpha}-\theta),}\\
\hspace*{-7mm}\lefteqn{\left[\frac{1}{r}\frac{\partial \Phi}{\partial\theta}-\frac{\partial \Psi_z}{\partial r}\right]_{r=\alpha,\theta}=U_{\alpha}\sin( \theta _{\alpha}-\theta)+\alpha \Theta _{\alpha},}\\
\hspace*{-7mm}\lefteqn{\left[   \frac{1}{r}\frac{d(r\Psi_{\theta})}{dr}\right]_{r=\alpha,\theta}=T_{\alpha}.}
\end{eqnarray}
\end{subequations}
Now, we are able to determine all  unknown constants involved in Eqs. (\ref{solutions general form}).

Owing to the linearity of the problem, general solution for the displacement field inside the cladding $\Omega$ can be written as $\bm{u}=\bm{u}^{(p)}+\bm{u}^{(a)}+\bm{u}^{(r)}$, where  $\bm{u}^{(p)}$  refers to the translation in $xy$ plane, $\bm{u}^{(a)}$  to the axial displacement in $\bm{e}_z$ direction, and $\bm{u}^{(r)}$	 to the rotation around the axis $(O,\bm{e}_z)$.  The first displacement field $\bm{u}^{(p)}$ can be expressed as 
\begin{eqnarray}\label{expression up cylinder}
\hspace*{-7mm}\bm{u}^{(p)}&=& A_1C_1\left\lbrace \frac{\mathrm{J}_1(hr)}{r}\bm{e}_x-h\mathrm{J}_2(hr)\cos\theta \bm{e}_r\right\rbrace + B_1C_1\left\lbrace \frac{\mathrm{Y}_1(hr)}{r}\bm{e}_x -h\mathrm{Y}_2(hr)\cos\theta\bm{e}_r\right\rbrace\notag \\ 
&+& E_1H_1\left\lbrace \frac{\mathrm{J}_1(\kappa r)}{r}\bm{e}_x+\kappa \mathrm{J}_2(\kappa r)\sin\theta\bm{e}_\theta\right\rbrace + F_1H_1\left\lbrace \frac{ \mathrm{Y}_1(\kappa r)}{r}\bm{e}_x+\kappa \mathrm{Y}_2(\kappa r)\sin\theta\bm{e}_\theta\right\rbrace  \notag \\
&+&A_1D_1\left\lbrace \frac{\mathrm{J}_1(hr)}{r}\bm{e}_y-h\mathrm{J}_2(hr)\sin\theta \bm{e}_r\right\rbrace + B_1D_1\left\lbrace \frac{\mathrm{Y}_1(hr)}{r}\bm{e}_y -h\mathrm{Y}_2(hr)\sin\theta\bm{e}_r\right\rbrace   \notag  \\ 
&+&E_1G_1\left\lbrace -\frac{\mathrm{J}_1(\kappa r)}{r}\bm{e}_y+\kappa \mathrm{J}_2(\kappa r)\cos\theta\bm{e}_\theta\right\rbrace + F_1G_1\left\lbrace -\frac{ \mathrm{Y}_1(\kappa r)}{r}\bm{e}_y+\kappa \mathrm{Y}_2(\kappa r)\cos\theta\bm{e}_\theta\right\rbrace 
\end{eqnarray}
where the first four terms, and the last four terms describe the displacements generated by the translations along $x$-axis and $y$-axis, respectively.  The unknown coefficients  in the above expression are the solutions to the linear systems  \eqref{linear system plane translation 1} and \eqref{linear system plane translation 2} given in  Appendix \ref{appendix expressions}. As such, we find in different terms, the solution of the plane translation given in Ref. \cite{sheng2005}  corresponding to the two DOFs related to translational motions. 

The displacement due to the translation along the cylinders' axis, i.e. $\bm{e}_{z}$, is expressed as
\begin{equation}
\bm{u}^{(a)}= \kappa\left\lbrace A\mathrm{J}_0(\kappa r)+B\mathrm{Y}_0(\kappa r) \right\rbrace \bm{e}_z.
\end{equation}
Finally the displacements related to the rotational motions are provided by  
\begin{equation}
\bm{u}^{(r)}=  \kappa \left\lbrace E_0G_0\mathrm{J}_1(\kappa r)+ F_0G_0\mathrm{Y}_1(\kappa r) \right\rbrace \bm{e}_{\theta}.
\end{equation}
After applying the boundary conditions \eqref{BC cylinder}, we obtain explicitly \footnote{In the limit $a\rightarrow 0$, i.e.,  making disappear the rigid core, the results presented in Ref. \cite{bonnet2017} can be easily obtained as a particular case of these general expressions.}
\begin{eqnarray}
\hspace*{-5mm}\bm{u}^{(a)}&=&
\begin{pmatrix}
\mathrm{J}_0(\kappa r) & \mathrm{Y}_0(\kappa r) \\ 
\end{pmatrix}
\begin{pmatrix}
\mathrm{J}_0(\kappa a) & \mathrm{Y}_0(\kappa a) \\ 
\mathrm{J}_0(\kappa b) & \mathrm{Y}_0(\kappa b) \\ 
\end{pmatrix}%
^{-1} 
\begin{pmatrix}
T_a \\ 
T_b \\ 
\end{pmatrix}%
\bm{e}_z  \notag \\
&=&%
\begin{pmatrix}
\displaystyle\frac{\mathrm{J}_0(\kappa r)\mathrm{Y}_0(\kappa b)-\mathrm{J}_0(\kappa b)\mathrm{Y}_0(\kappa r)}{%
\mathrm{J}_0(\kappa a)\mathrm{Y}_0(\kappa b)-\mathrm{J}_0(\kappa b)\mathrm{Y}_0(\kappa a)} & \displaystyle\frac{%
\mathrm{J}_0(\kappa a)\mathrm{Y}_0(\kappa r)-\mathrm{J}_0(\kappa r)\mathrm{Y}_0(\kappa a)}{\mathrm{J}_0(\kappa
a)\mathrm{Y}_0(\kappa b)-\mathrm{J}_0(\kappa b)\mathrm{Y}_0(\kappa a)} \\ 

\end{pmatrix}
\begin{pmatrix}
T_a \\ 
T_b
\end{pmatrix}
\bm{e}_z,  \label{eq:ua}
\end{eqnarray}

\begin{eqnarray}
\hspace*{-5mm}\bm{u}^{(r)}&=& 
\begin{pmatrix}
\mathrm{J}_1(\kappa r) & \mathrm{Y}_1(\kappa r) \\ 
\end{pmatrix}
\begin{pmatrix}
\mathrm{J}_1(\kappa a) & \mathrm{Y}_1(\kappa a) \\ 
\mathrm{J}_1(\kappa b) & \mathrm{Y}_1(\kappa b) \\ 
\end{pmatrix}%
^{-1}
\begin{pmatrix}
a\Theta_a \\ 
b\Theta_b \\ 
\end{pmatrix}%
\bm{e}_\theta  \notag \\
&=& 
\begin{pmatrix}
\displaystyle a\frac{\mathrm{J}_1(\kappa r)\mathrm{Y}_1(\kappa b)-\mathrm{J}_1(\kappa b)\mathrm{Y}_1(\kappa r)}{%
\mathrm{J}_1(\kappa a)\mathrm{Y}_1(\kappa b)-\mathrm{J}_1(\kappa b)\mathrm{Y}_1(\kappa a )} & \displaystyle b%
\frac{\mathrm{J}_1(\kappa a)\mathrm{Y}_1(\kappa r)-\mathrm{J}_1(\kappa r)\mathrm{Y}_1(\kappa a)}{\mathrm{J}_1(\kappa
a)\mathrm{Y}_1(\kappa b)-\mathrm{J}_1(\kappa b)\mathrm{Y}_1(\kappa a )} \\ 
\end{pmatrix}
\begin{pmatrix}
\Theta_a \\ 
\Theta_b%
\end{pmatrix}%
\bm{e}_\theta  \label{eq:ur}
\end{eqnarray}

It is important to note that,  these axial and rotational contributions of the motion involves only $\kappa$ (and not $h$). This is due to the state of shear dictating such  kinematics. This constitutes a key point in order to achieve an anisotropic effective density, and obtain different characteristics for different directions. 

We now use the equations of motion and the stresses that are known in the cladding, to deduce the relation between the displacements of the core cylinder and the embedding matrix. The equations of motion are written  as
\begin{subequations} \label{eq:PFD}
\begin{eqnarray}
-M_a\omega^2\bm{V}_a=\int\limits_{\Gamma_a}\bm{\sigma}.\bm{n}\,dS  \\ 
 -I_a\omega^2 \bm{\Theta}_a=\int\limits_{\Gamma_a}\bm{r}\times (\bm{\sigma}.\bm{n})\,dS 
\end{eqnarray}
\end{subequations}
where  
$M_a=\rho_a(\pi a^2 L)$  is the mass of the core cylinder with the density $\rho_{a}$, and $I_a=M_a a^2/2$ is its moment of inertia.  The stress tensor $\bm{\sigma}=\mu(\bm{\nabla}\bm{u}+\bm{\nabla}\bm{u}^T)+\lambda (\bm{\nabla}.\bm{u})\bm{\mathcal{I}}$,  with $\bm{u}^T$ the transpose of $\bm{u}$ and $\bm{\mathcal{I}}$ the identity matrix,  is known since the displacement $\bm{u}$ in the cladding is fully determined.  After some simple calculations, we obtain the relationship between displacements of the core and the matrix, as follows 
\begin{equation}\label{displacements}
\begin{cases}
\;\displaystyle U_a=\frac{-\gamma g^{(p)}(\omega)}{R^{(p)}(\omega)-\rho_a/\rho}\; U_b := H^{(p)}(\omega)\; U_b \\ 
\\ 
\;\displaystyle T_a=\frac{-g^{(a)}(\omega)}{R^{(a)}(\omega)-\rho_a/\rho}\; T_b := H^{(a)}(\omega)\; T_b \\ 
\\ 
\;\displaystyle \Theta_a=\frac{-\gamma g^{(r)}(\omega)}{R^{(r)}(\omega)-\rho_a/\rho}\; \Theta_b := H^{(r)}(\omega)\; \Theta_b \\ 
\end{cases}
\end{equation}
where $\gamma:= b/a$. The expressions of $R^{(p)}(\omega)$ and $g^{(p)}(\omega)$, involved in plane translational motion are given in the Appendix \ref{appendix expressions} [Eqs. \eqref{gp and Rp}], which can also be found also in Ref. \cite{sheng2005}. The other components related to axial translation and rotation are expressed as   
\begin{subequations}
\begin{eqnarray}
\hspace*{-7mm}\begin{cases}
\;\displaystyle g^{(a)}(\omega)=\frac{2}{\kappa a}\frac{\mathrm{J}_0(\kappa a)\mathrm{Y}_1(\kappa a)-\mathrm{J}_1(\kappa a)\mathrm{Y}_0(\kappa a)}{\mathrm{J}_0(\kappa a)\mathrm{Y}_0(\kappa b)-\mathrm{J}_0(\kappa b)\mathrm{Y}_0(\kappa a)} \\ 
\\ 
\;\displaystyle R^{(a)}(\omega)=\frac{2}{\kappa a}\frac{\mathrm{J}_1(\kappa a)\mathrm{Y}_0(\kappa b)-\mathrm{J}_0(\kappa b)\mathrm{Y}_1(\kappa a)}{\mathrm{J}_0(\kappa a)\mathrm{Y}_0(\kappa b)-\mathrm{J}_0(\kappa b)\mathrm{Y}_0(\kappa a)} \\ 
\end{cases}\\  
\hspace*{-7mm}\begin{cases}
\;\displaystyle g^{(r)}(\omega)=\frac{4}{\kappa a}\frac{\mathrm{J}_1(\kappa a)\mathrm{Y}_2(\kappa a)-\mathrm{J}_2(\kappa a)\mathrm{Y}_1(\kappa a)}{\mathrm{J}_1(\kappa a)\mathrm{Y}_1(\kappa b)-\mathrm{J}_1(\kappa b)\mathrm{Y}_1(\kappa a)} \\ 
\\ 
\;\displaystyle R^{(r)}(\omega)=\frac{4}{\kappa a}\frac{\mathrm{J}_2(\kappa a)\mathrm{Y}_1(\kappa b)-\mathrm{J}_1(\kappa b)\mathrm{Y}_2(\kappa a)}{\mathrm{J}_1(\kappa a)\mathrm{Y}_1(\kappa b)-\mathrm{J}_1(\kappa b)\mathrm{Y}_1(\kappa a)}\\ 
\end{cases}%
\end{eqnarray}
\end{subequations}

\begin{table}
\caption{\label{table}%
Material properties of the three-component unit-cell.
}
\begin{ruledtabular}
\begin{tabular}{cccc}
Material & Epoxy (matrix) & Silicone (cladding) & Lead (core) \\ 
$\lambda$ [$\rm Pa$] & $4.43 \times10^9$ & $6\times10^{5}$ & $4.23\times10^{10}$ \\ 
$\mu$ [$\rm Pa$] & $1.59\times10^9$ & $4\times10^4$ & $1.49\times10^{10}$ \\ 
$\rho$ [$\rm kg.m^{-3}$] & $1.18\times10^3$ & $1.3\times10^3$ & $11.6\times10^3$ \\ 
\end{tabular}
\end{ruledtabular}
\end{table}

\begin{figure}
\hspace*{-6mm}\includegraphics[height=8 cm,draft=false]{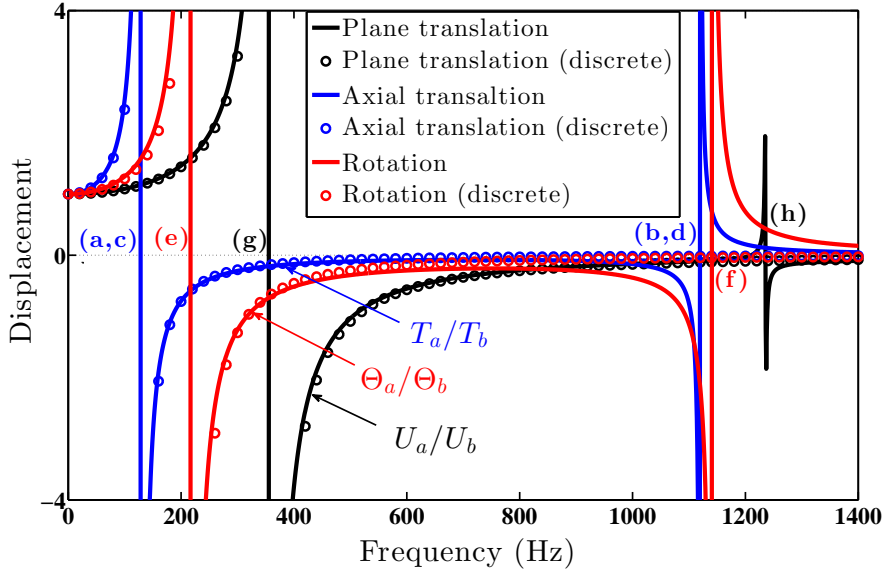} %
\caption{Displacement of the cylindrical core normalized to and exited by that of the matrix, based on the continuum model and discrete mass-spring model, and related to three modes: plane translation, axial translation, and rotation. The displacement field patterns inside the elastic coating layer (between the core and the matrix) for the frequency points (a), (b),...,(h) will be shown in Fig. \ref{fig:fields}.}
\label{fig:Amp_2D}
\end{figure}

\begin{figure}
\hspace*{-2mm}\includegraphics[height=2.75 cm,draft=false]{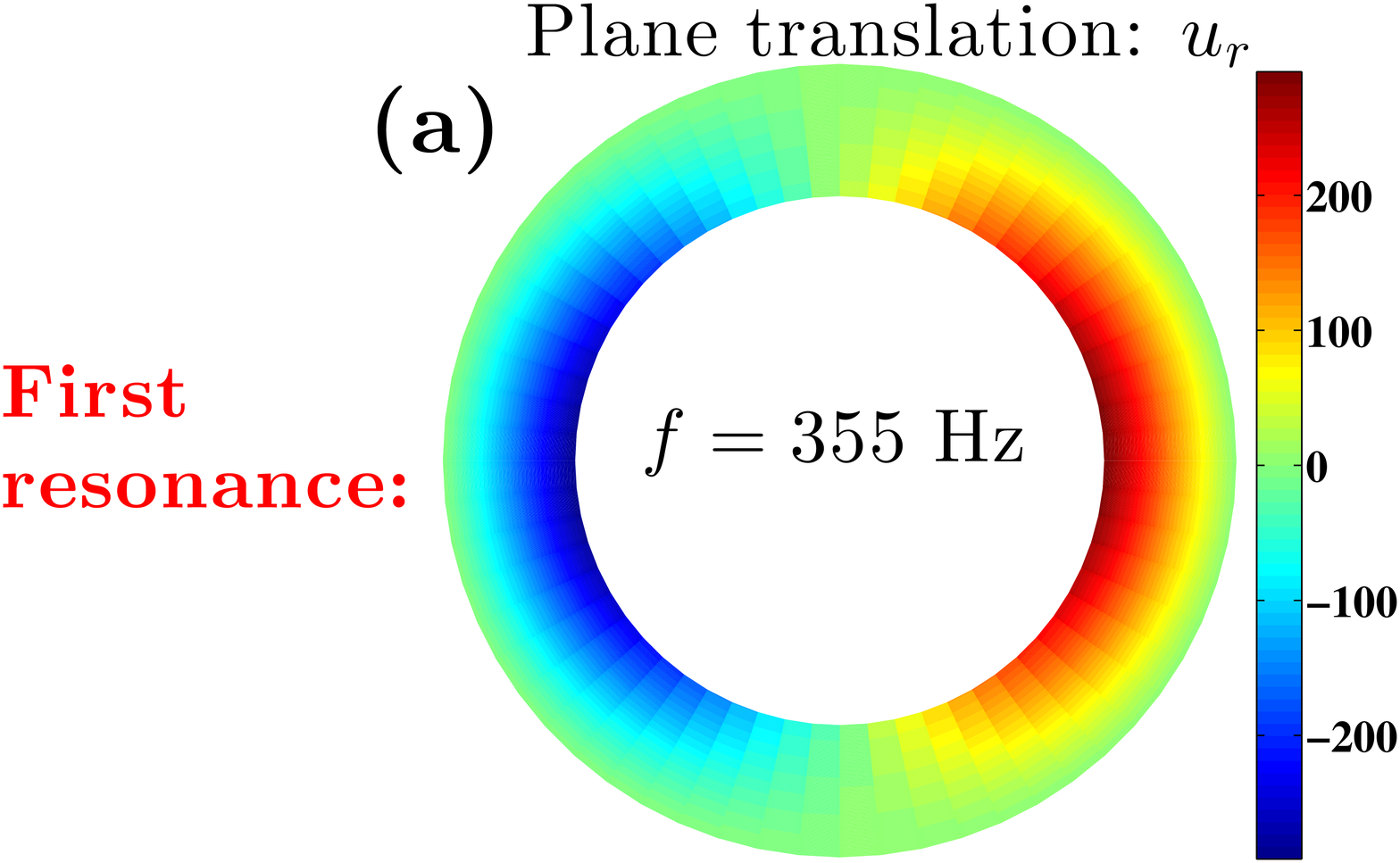}
\hspace*{-8.5mm}\includegraphics[height=2.75 cm,draft=false]{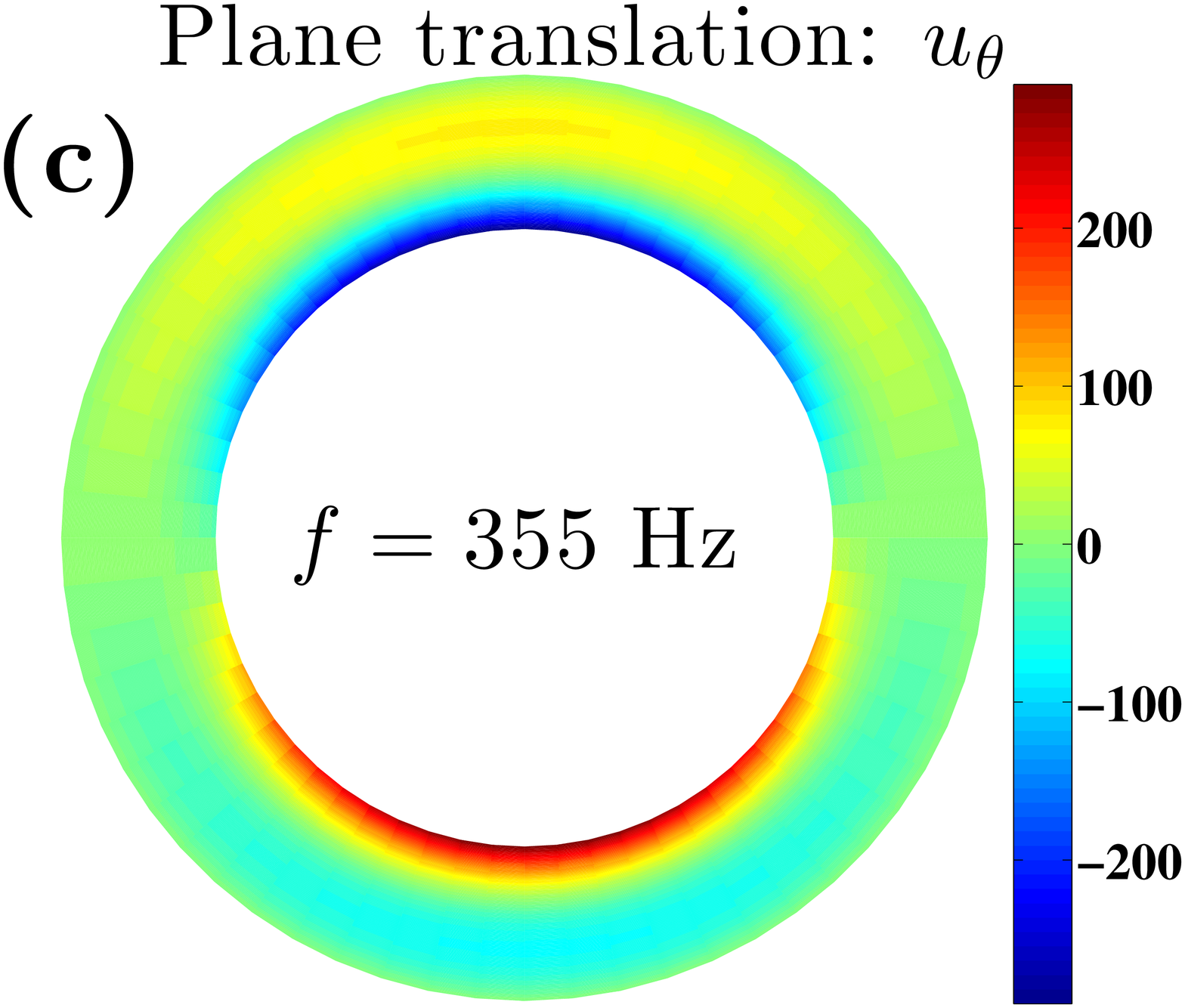}
\hspace*{-8.5mm}\includegraphics[height=2.75 cm,draft=false]{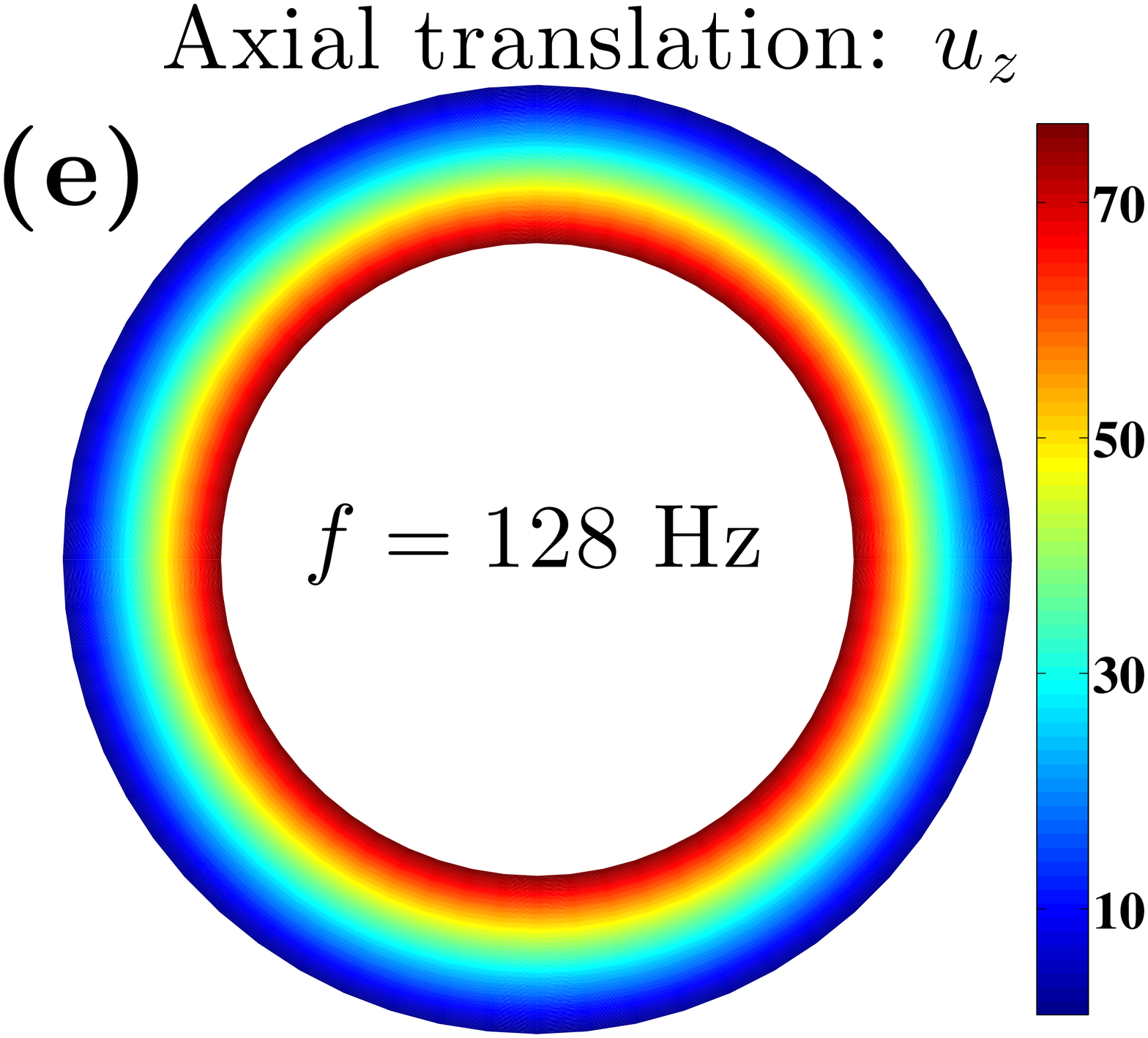}
\hspace*{-8.5mm}\includegraphics[height=2.75 cm,draft=false]{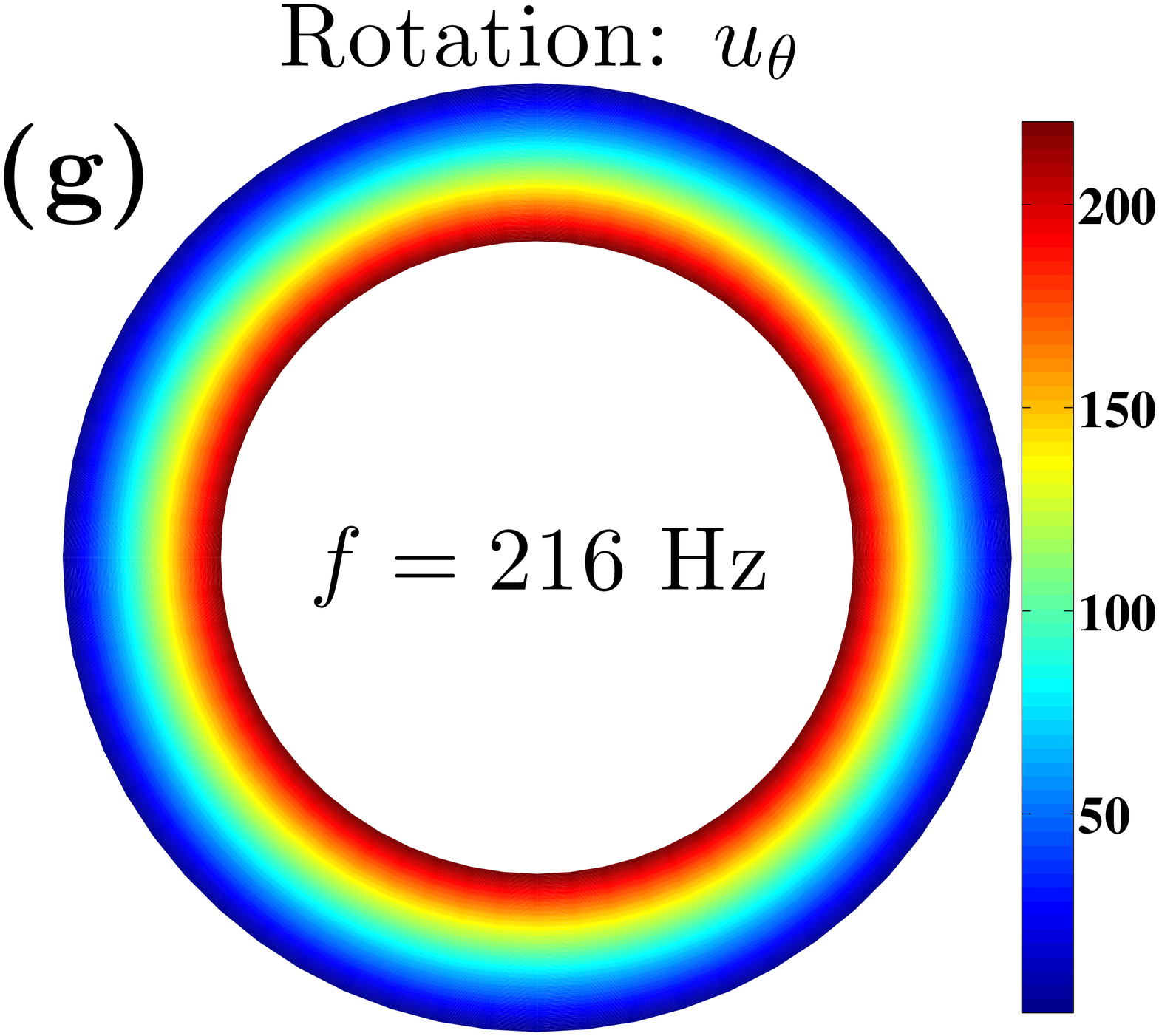} \\
\hspace*{-2mm}\includegraphics[height=2.75 cm,draft=false]{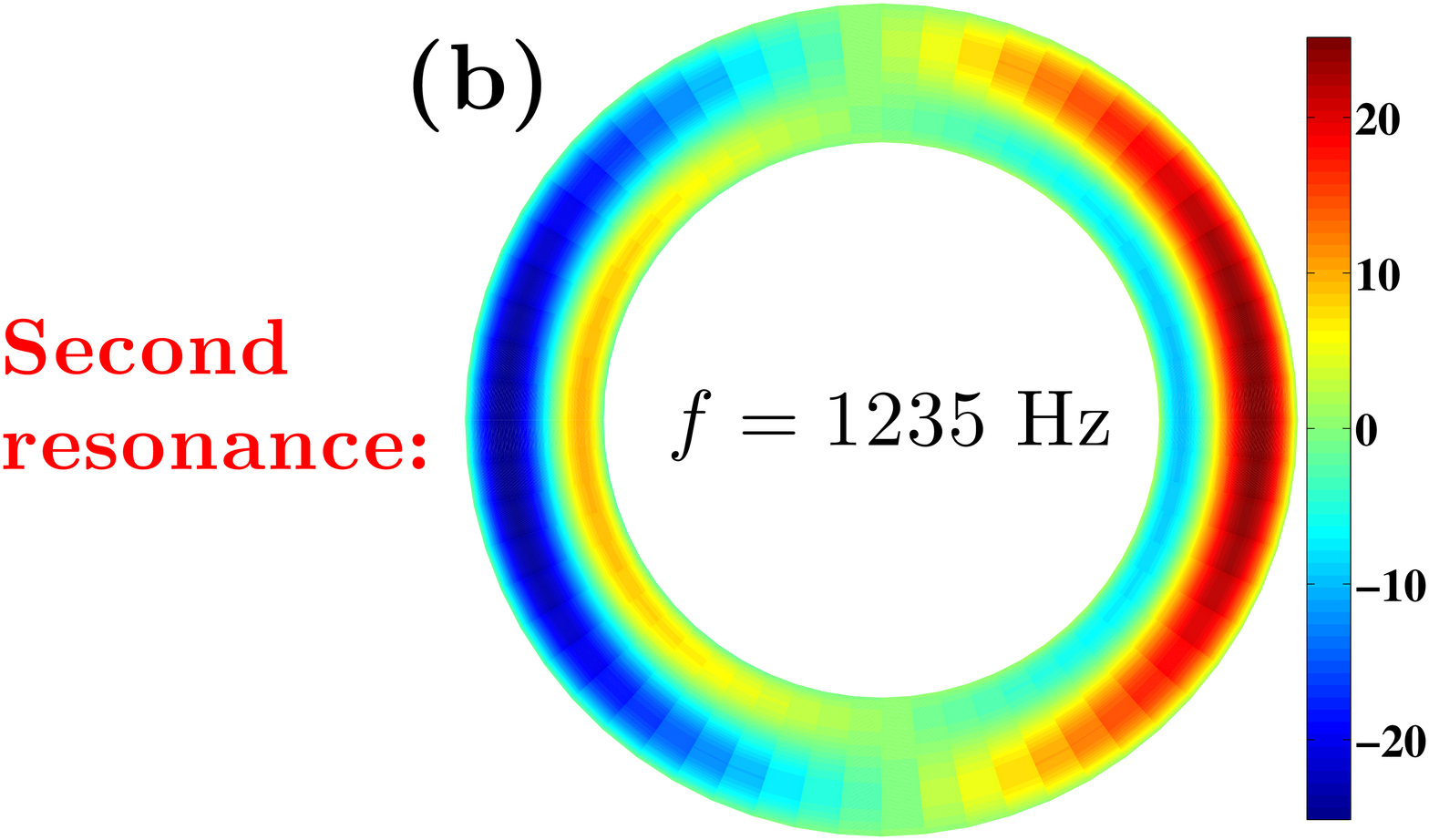}
\hspace*{-8.5mm}\includegraphics[height=2.75 cm,draft=false]{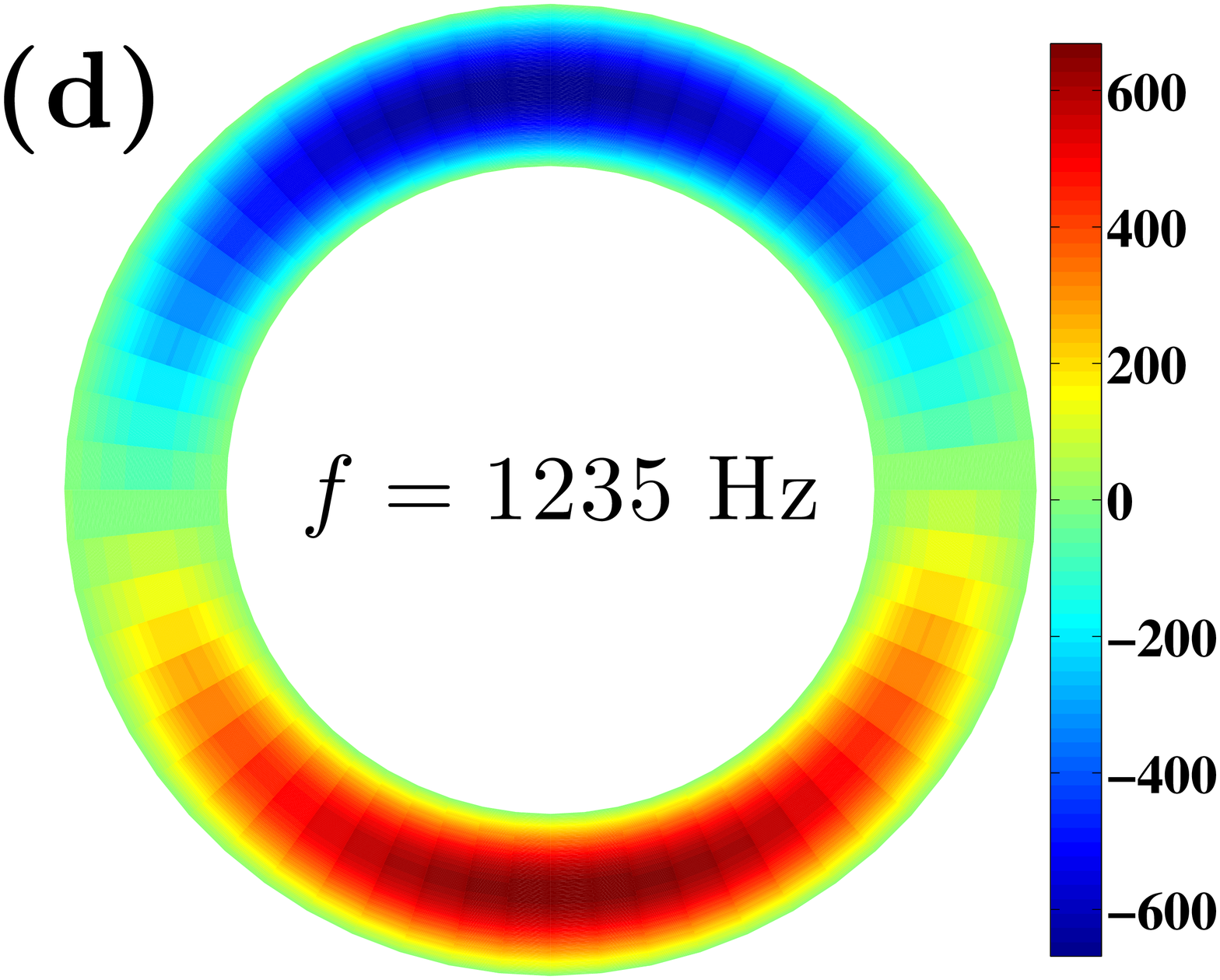}
\hspace*{-8.5mm}\includegraphics[height=2.75 cm,draft=false]{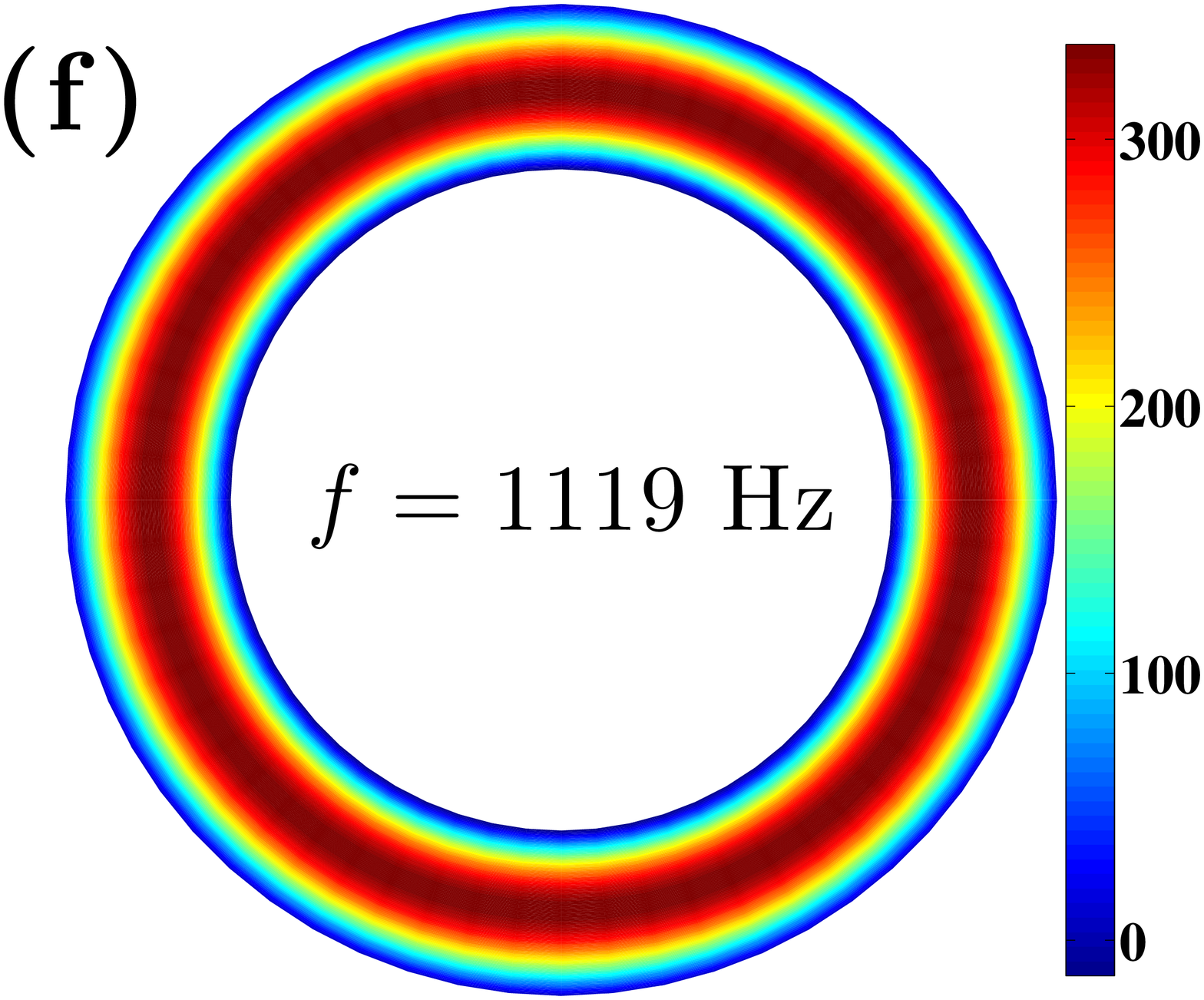}
\hspace*{-8.5mm}\includegraphics[height=2.75 cm,draft=false]{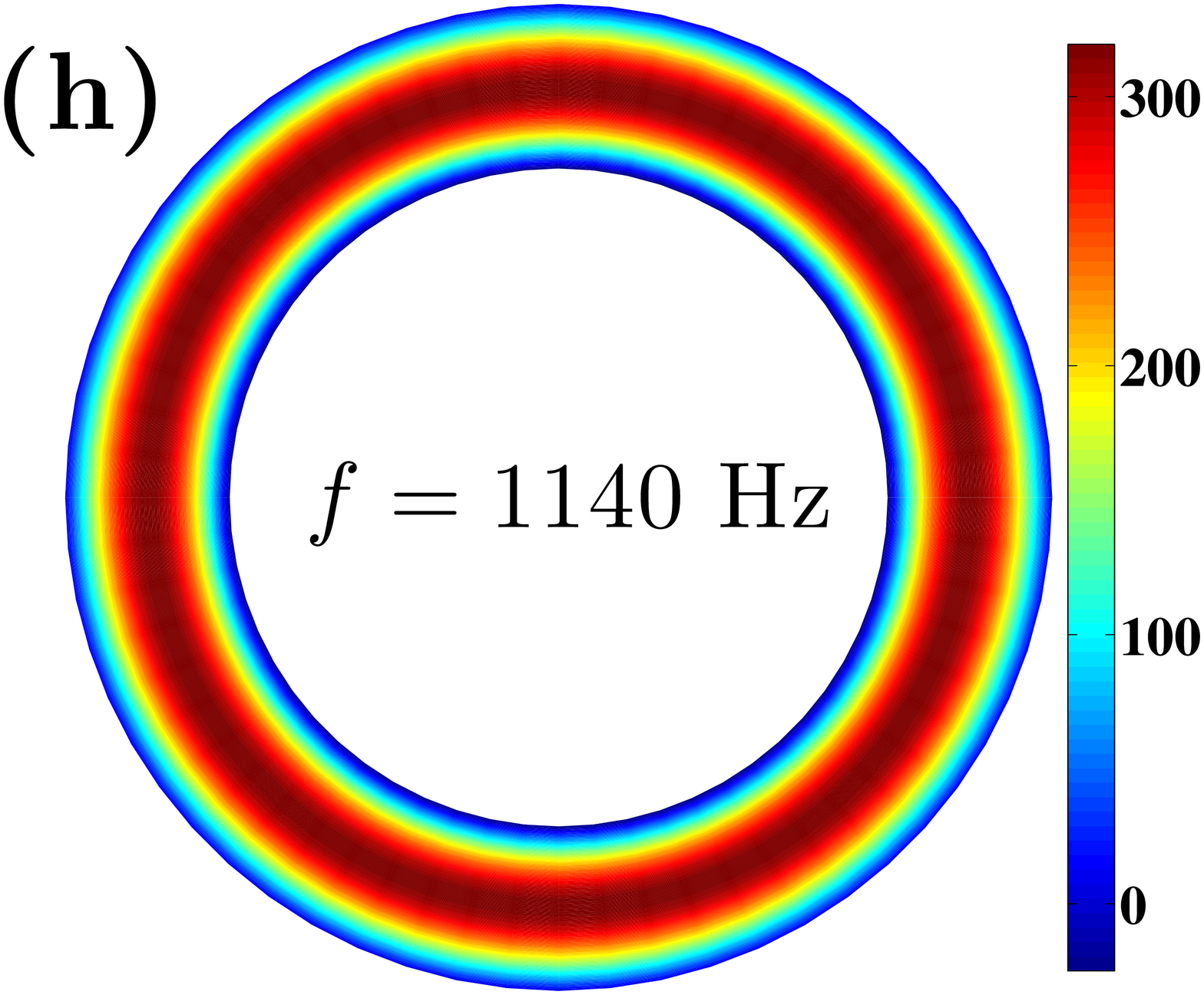}
\caption{Displacement field distributions inside the coating layer close to the first and second resonance frequencies for the modes related to plane translation, axial translation, and rotation. The displacements are normalized such that they have fixed unit value on the outer boundary of the coating, i.e., at $r=b$ in Fig. \ref{fig:CL}. The field patterns in (a), (b), ..., (h) are related to respective frequency points shown in Fig. \ref{fig:Amp_2D}.}
\label{fig:fields}
\end{figure}

Fig. \ref{fig:Amp_2D} shows the evolution of the displacement of the core cylinder for the three modes in function of frequency, represented by $H^{(p)}(\omega)$, $H^{(a)}(\omega)$ and $H^{(r)}(\omega)$, using the same material properties as in Ref.\cite{sheng2000}, that are shown in Table \ref{table} and geometrical parameters with  the filling fraction for coated cylinders being 40\%, 5.0 mm for the radius of the lead cylinder, and the coating thickness is 2.5 mm. For these modes of plane translation, axial translation, and rotation, the displacement field patterns inside the cladding close to the first and second resonance frequencies are depicted in Fig \ref{fig:fields}. Through Fig. \ref{fig:fields}, we observe that the field amplitudes are maximum on the boundary of the rigid cylinder at the first resonance frequency, while it is maximum inside the cladding at the second resonance frequency. The field patterns in Fig. \ref{fig:fields}a to Fig. \ref{fig:fields}h correspond to the frequency points (a) to (h) in Fig. \ref{fig:Amp_2D}. Behaviors of $H^{(a)}(\omega)$ and $H^{(r)}(\omega)$ are similar to that of $H^{(p)}(\omega)$, although obviously the resonances relating to different types of motions, when $R(\omega)\equiv\rho_a/\rho$, are located at different frequencies.  Indeed, the first resonances for axial, rotational and plane modes are obtained, respectively,  at $f_0^{(a)}\approx 128$ Hz, $f_0^{(r)}\approx 217$ Hz and $f_0^{(p)}\approx 355$ Hz. These resonances concern rigid-body mode insofar as the core vibrates as a whole within the elastic cladding. In this case, the dynamics of the structural unit suggests that the unit can be reasonably assimilated to a simple mass-spring system for the translational \cite{huang2009} as well as rotational mode \cite{bonnet2015}, by taking the hard cylinder as the mass, the cladding as the spring, which is attached to the mass on one side and to the rigid matrix on its other side. The mass-spring system is exited by the motion of the matrix.

Based on this mass-spring model the first resonance frequencies $f_0^{(p)}$,  $f_0^{(a)}$, and $f_0^{(p)}$ for plane translation, axial translation, and rotation motions, respectively, can be obtained as follows  
\begin{eqnarray}
\hspace*{-7mm}\lefteqn{\displaystyle  f_0^{(p)}=f_0\left\lbrace  \frac{8(1-\nu)(3-4\nu)\gamma^2}{(3-4\nu)^2\ln\gamma-\frac{\gamma^2-1}{\gamma^2+1}}\right\rbrace^{1/2},
\;\displaystyle f_0^{(a)}=f_0\left( \frac{2\gamma^2}{\ln\gamma}  \right)  ^{1/2},\;  \displaystyle f_0^{(r)}=f_0\left(  \frac{8\gamma^4}{\gamma^2-1}\right)^{1/2},} 
\label{eq:freq_2D}
\end{eqnarray}
where $f_0=(1/2\pi)\sqrt{\mu/\rho_a b^2}$. With our present configuration, the above formulas give $f_0^{(a)}\approx 131$ Hz, $ f_0^{(r)}\approx 224$ Hz, and $f_0^{(p)}\approx 360$ Hz. The good agreement with the resonance frequencies from the continuum model that is observed  allows us a very fast and simple estimation of the first resonance frequencies, especially in terms of geometrical and material parameters associated with each constituent. To obtain the displacement based on the mass-spring model, we need to calculate the effective spring constants for each of the translational and rotational modes. The corresponding expressions for the effective spring constants in terms of microscale material parameters, relating to the three modes, are found in Appendix \ref{appendix stiffness}.  

\begin{figure}
\hspace*{-10mm}\includegraphics[height=8 cm,draft=false]{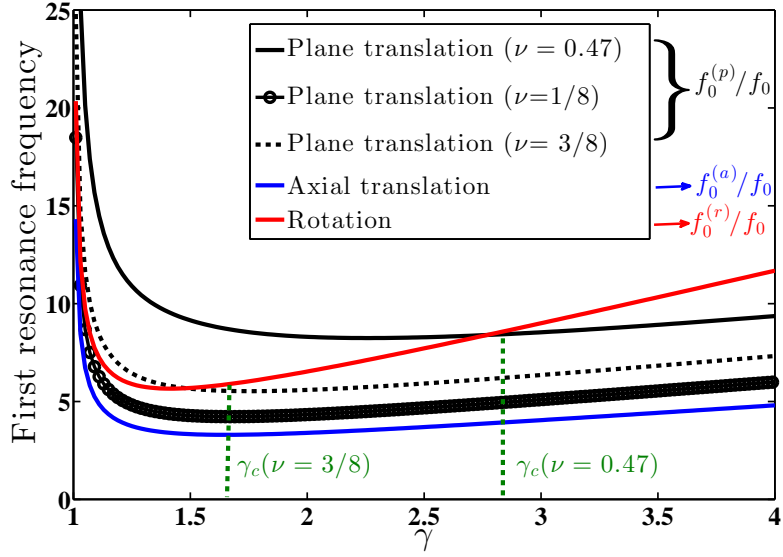} %
\caption{Evolution of the first resonance frequencies normalized by $f_0$ and estimated by the discrete mass-spring model [Eq. \eqref{eq:freq_2D}], with the form factor $\gamma=b/a$. The dependence of plane-translation mode on the Poisson's ratio $\nu$ of the coating  is also illustrated. The Poisson's ratio takes two representative values $1/8$ and $3/8$ in the intervals $[0, 1/4]$ and $[1/4, 1/2]$, respectively. The value of $\nu=0.47$ corresponds to the material parameters studied as an example throughout this paper (see Table \ref{table}). The axial-translation and rotation modes are independent of $\nu$.}
\label{fig:freq_2D}
\end{figure}

The study of relations \eqref{eq:freq_2D} makes it possible to demonstrate that $f_0^{(a)}$ always remains smaller than the other two resonance frequencies, whatever the geometrical parameters and material properties of the coating layer. This can be observed in Fig. \ref{fig:freq_2D}.  The relations \eqref{eq:freq_2D} also show that unlike the Poisson's ratio $\nu = \lambda/\left\lbrace 2(\lambda + \mu)\right\rbrace$, the Lamé coefficient $\mu$ does not change the positions of the frequencies $f_0^{(p)}$, $f_0^{(a)}$, and $f_0^{(r)}$ relative to each other. More precisely, for Poisson's ratio laying in the interval $0 \leq \nu \leq 1/4$,  the rotational resonance  always manifests itself at higher frequency compared with the plane translation resonance ($f_0^{(p)}<f_0^{(r)}$), whereas for Poisson's ratio $1/4 <\nu \leq 1/2$ the relative position of these two frequencies evolves with the form factor $\gamma = b/a $. This implies that for small inclusions such that $\gamma_c (\nu)<\gamma$, we have  $ f_0^{ (p)}<f_0^{(r)}$, and vice versa otherwise; where $\gamma_c(\nu)$ is the unique solution to the equation $f_0^{(p)}(\gamma)=f_0^{(r)}(\gamma)$, i.e., $\ln\gamma=\dfrac{1}{(3-4\nu)^2}\left\lbrace (1-\nu)(3-4\nu)\left(1-\dfrac{1}{\gamma^2}\right)+1-\dfrac{2}{\gamma^2+1}\right\rbrace $. We note that $\gamma_c (\nu)$ is an increasing function, varying from $1$ for $\nu = 1/4$, to about $3.80 $ when $\nu = 1/2$. Fig. \ref{fig:freq_2D} illustrates these remarks; in particular it shows the evolution of $f_0^{(p)}$ with $\gamma$ for three representative values of $\nu$: $1/8$ ($0 \leq \nu \leq 1/4$), $3/8$ ($1/4 <\nu \leq 1/2$), and $0.47$ that is the value for the structure studied in particular in this paper (Table \ref{table}),  laying also in the interval $1/4 <\nu \leq 1/2$.

Although, as seen in Fig. \ref{fig:Amp_2D}, the mass-spring model describes well the dynamics of the material at lower frequencies including the first resonances, it fails to predict the second resonance behavior for all of the three modes. This is due to the fact that the displacement field at the second resonance which is mainly produced by the standing waves inside the coating layer, while the lead cylinder follows the motion of the matrix and remains practically at rest (see Ref. \cite{sheng2005} for plane translation mode). We know that the spring, modeling the coating layer, has been assumed to be massless, and consequently this discrete model by its construction cannot describe the second resonance. That being said, if we consider a more complex discrete model by taking into account a mass for the mode-related springs a second local resonance in the elastic medium will appear. However, the outcome of latter model does not correspond to that of continuum model; particularly because in the mass-spring model the springs function in one dimension, while at the second local resonance, the displacement fields in the coating layer can be distributed in three dimensions taking, e.g. the shear effects into account. The distribution of displacement field in the $xy$ plane for each of the three modes is shown in Fig. \ref{fig:fields}, close to the first and to the second resonance frequencies. In this figure, comparison of the field pattern at the first resonance for either of the modes with those arising from the same mode but at the second resonance, shows that at the second resonance the displacement field is distributed in space to a significantly greater extent than that related to the first resonance. In addition, the evaluation of the spring constants is performed in a quasi-static regime, which is further from the frequency-dependent nature of the second resonance occurring inside the elastic layer at higher frequencies.  The resonance frequencies in Figs. \ref{fig:fields}a to \ref{fig:fields}h, correspond to the points (a), (b), ..., and (h) in Fig. \ref{fig:Amp_2D}. The second resonance of the structure almost coincides with the first cladding resonance. This first cladding frequency can be estimated by searching for the first root of the determinant, which is a classical technique for modal analysis. In fact, we can proceed to find the solutions, by imposing a zero-displacement for the lead cylinder within the continuum model, assuming that the resonance occur only inside the elastic medium. Thus, we are led to determine the first root of $\Delta^{(p)}$, $\Delta^{(a)}$, and $\Delta^{(r)}$ in order to find the second resonance frequency of the plane translation, axial translation, and rotation mode, respectively, through the following equations  
\begin{eqnarray}\label{eq:determinant}
\hspace*{-7mm}\Delta^{(p)}(\omega)&:= &  \det \mathcal{D}=\det\begin{pmatrix}
[E(ha)] & [F(\kappa a)] \\
[E(hb)] & [F(\kappa b)] \\
\end{pmatrix}=0,\notag \\
\hspace*{-7mm}\Delta^{(a)}(\omega)&:= & \mathrm{J}_0(\kappa a)\mathrm{Y}_0(\kappa b)-\mathrm{J}_0(\kappa b)\mathrm{Y}_0(\kappa a)=0,\\
\hspace*{-7mm}\Delta^{(r)}(\omega)&:= &\mathrm{J}_1(\kappa a)\mathrm{Y}_1(\kappa b)-\mathrm{J}_1(\kappa b)\mathrm{Y}_1(\kappa a)=0.\notag
\end{eqnarray}
We refer to the equation \eqref{linear system plane translation 1} in Appendix \ref{appendix expressions} for the derivation of the first relation and the definition of the associated block matrix $\mathcal{D}$. The two last relations arise from the equations \eqref{eq:ua} and \eqref{eq:ur}.  The first and second local-resonance frequencies related to different modes,  and calculated by different methods, are collected in Table \ref{table resonance}, in order to easily compare these values associated with their respective method of calculation. In particular, it can be noticed  that the relations \eqref{eq:determinant} provide good estimations of the second resonance frequencies.      
\begin{table}
\caption{Fist resonance (R1) and second resonance (R2) frequencies obtained by the continuum model, discrete model [Eqs. \eqref{eq:freq_2D}], and estimation [Eqs. \eqref{eq:determinant}], for plane translation, axial translation, and rotation modes.}
\begin{ruledtabular}
\begin{tabular}{cccc}
Frequency & Plane (Hz) & Axial (Hz)& Rotation (Hz)\\ 
 \hline
Continuum model: R1& $355$ & $128$ & $217$ \\ 
Discrete model : R1& $360$ & $131$ & $224$ \\  
Continuum model: R2& $1236$ & $1120$ & $1140$ \\
Estimation: R2 & $1236$ & $1107$ & $1116$ \\ 
\end{tabular}
\end{ruledtabular}
\label{table resonance}
\end{table}

%

\section{Media with spherical inclusions} 
\label{section sphere}

Here, we study  the same kind of problem as in the previous section with the only difference that the geometry of the structural unit is changed: the cylinder of radius $a$ and height $L\gg a$ is replaced by a hard ball of radius $a$. An embedding matrix has now  spherical cavities of radius $b$, that are filled  with the hard spheres and a concentric elastic cladding such that $a< R< b$ (Fig. \ref{fig:coord}a). All these structural components are made of the same materials as in the case of cylindrical system  (see Table \ref{table}), so that the same simplifications can be applied to the corresponding kinematics. By virtue of the analysis in the previous section with coaxial cylinders, it is evident to claim, without any calculations,  that for an arbitrary motion of the embedding matrix as a rigid body, the same type of motion is  induced for the embedded hard ball. The superposition principle, verified for the previous case, suggests us that translations and rotations are totally uncoupled. A translation in a fixed direction or a rotation around a fixed direction, will lead, respectively, to the same translation or rotation for the hard ball,  but with possibly different amplitude and/or phase. 

Moreover, assuming a total isotropy of the problem, it is now convenient to consider, independently, only one translation in a certain direction and one rotation around a fixed axis to describe the whole kinematics of this problem. The former problem regarding the translational mode based on the continuum model has been solved in Ref. \cite{sheng2005}. In fact, regarding the translational mode, we only need to replace the Bessel functions in the cylindrical system by spherical Bessel functions and use the formalism explained in Appendix \ref{appendix expressions}. The aim of this section is to describe the kinematics of the rotational mode according to the continuum model, and compare with its counterpart built upon the discrete mass-spring model. Related equations of motion to be solved, for the cladding zone and its boundaries are stated as follows 
\begin{subequations}\label{eq:GenePb_2}
\begin{eqnarray} 
\hspace*{-7mm}\lefteqn{\;\displaystyle (\lambda+2\mu)\bm{\nabla}({\bm{\nabla}}.\bm{u})-\mu\bm{\nabla}\times(\bm{\nabla}\times \bm{u})=-\rho\omega^2\bm{u},} \\ \label{equation motion sphere 1}
\hspace*{-7mm}\lefteqn{\;\displaystyle \bm{u}|_{R=\alpha}=\bm{\Theta}_\alpha\times\bm{r}|_{R=\alpha}\ ,\ \ \alpha=a,b.} \label{equation motion sphere 2}
\end{eqnarray}
\end{subequations}
Without loss of generality, we can choose $\bm{\Theta}_\alpha=\Theta_\alpha\bm{e}_z$ (see Fig. \ref{fig:coord}b for the coordinate system), so that the solution is of the form $\bm{u}=u_\theta(R,\varphi)\bm{e}_\theta\equiv f(R)\sin\varphi \bm{e}_\theta$. The equation \eqref{equation motion sphere 1} becomes $\displaystyle f^{\prime \prime}(R)+(2/R)f^{\prime }(R)+(\kappa^2- 2/R^2 )f(R)=0$, with the boundary conditions $f(R=\alpha)=\alpha\Theta_{\alpha}$, following \eqref{equation motion sphere 2}. We can find  the solution explicitly in terms of the first order spherical Bessel functions of the first -$\mathrm{j}_1$- and second -$\mathrm{y}_1$- kind : 
\begin{eqnarray}
\hspace*{-7mm}\bm{u}&=& 
\begin{pmatrix}
\mathrm{j}_1(\kappa R) & \mathrm{y}_1(\kappa R) \\ 
\end{pmatrix}
\begin{pmatrix}
\mathrm{j}_1(\kappa a) & \mathrm{y}_1(\kappa a) \\ 
\mathrm{j}_1(\kappa b) & \mathrm{y}_1(\kappa b) \\ 
\end{pmatrix}%
^{-1} 
\begin{pmatrix}
a\Theta_a \\ 
b\Theta_b \\ 
\end{pmatrix}
\sin\varphi\bm{e}_\theta  \notag \\
\hspace*{-7mm}&=& 
\begin{pmatrix}
\displaystyle a\frac{\mathrm{j}_1(\kappa R)\mathrm{y}_1(\kappa b)-\mathrm{j}_1(\kappa b)\mathrm{y}_1(\kappa R)}{\mathrm{j}_1(\kappa a)\mathrm{y}_1(\kappa b)-\mathrm{j}_1(\kappa b)\mathrm{y}_1(\kappa a )} & 
\displaystyle b\frac{\mathrm{j}_1(\kappa a)\mathrm{y}_1(\kappa R)-\mathrm{j}_1(\kappa R)\mathrm{y}_1(\kappa a)}{\mathrm{j}_1(\kappa a)\mathrm{y}_1(\kappa b)-\mathrm{j}_1(\kappa b)\mathrm{y}_1(\kappa a )}
\\ 
\end{pmatrix}
\begin{pmatrix}
\Theta_a \\ 
\Theta_b
\end{pmatrix}
\sin\varphi\bm{e}_\theta  \label{eq:ur_3D}
\end{eqnarray}

As in Sec. \ref{section cylinder}, we  derive the relation between the displacement of the lead sphere and the embedding matrix by applying the equation of motion for the coated sphere, and using the fact that the stresses in the cladding can be known through the displacements [Eq. \eqref{eq:ur_3D}]: 
\begin{equation}
-I_a\omega^2\bm{\Theta}_a=\int\limits_{\Gamma_a}\bm{r}\times (\bm{\sigma}.\bm{n})\,dS  \label{eq:PFD_3D}
\end{equation}
where the moment of inertia of the hard sphere $I_a=2M_a a^2/5 $ around $\bm{e}_z$. We finally obtain
\begin{equation}
\Theta_a=\frac{-\gamma g_*^{(r)}(\omega)}{R_*^{(r)}(\omega)-\rho_a/\rho}\;\Theta_b := H_*^{(r)}(\omega)\;\Theta_b,
\end{equation}
provided that
\begin{equation}
\hspace*{-7mm}\begin{cases}
\;\displaystyle g_*^{(r)}(\omega)=\frac{5}{\kappa a}\frac{\mathrm{j}_1(\kappa a)\mathrm{y}_2(\kappa a)-\mathrm{j}_2(\kappa a)\mathrm{y}_1(\kappa a)}{\mathrm{j}_1(\kappa a)\mathrm{y}_1(\kappa b)-\mathrm{j}_1(\kappa b)\mathrm{y}_1(\kappa a)} \\ 
\\ 
\;\displaystyle R_*^{(r)}(\omega)=\frac{5}{\kappa a}\frac{\mathrm{j}_2(\kappa a)\mathrm{y}_1(\kappa b)-\mathrm{j}_1(\kappa b)\mathrm{y}_2(\kappa a)}{\mathrm{j}_1(\kappa a)\mathrm{y}_1(\kappa b)-\mathrm{j}_1(\kappa b)\mathrm{y}_1(\kappa a)} \\ 
\end{cases}
\end{equation}

\begin{figure}
\hspace*{-6mm}\includegraphics[height=8 cm,draft=false]{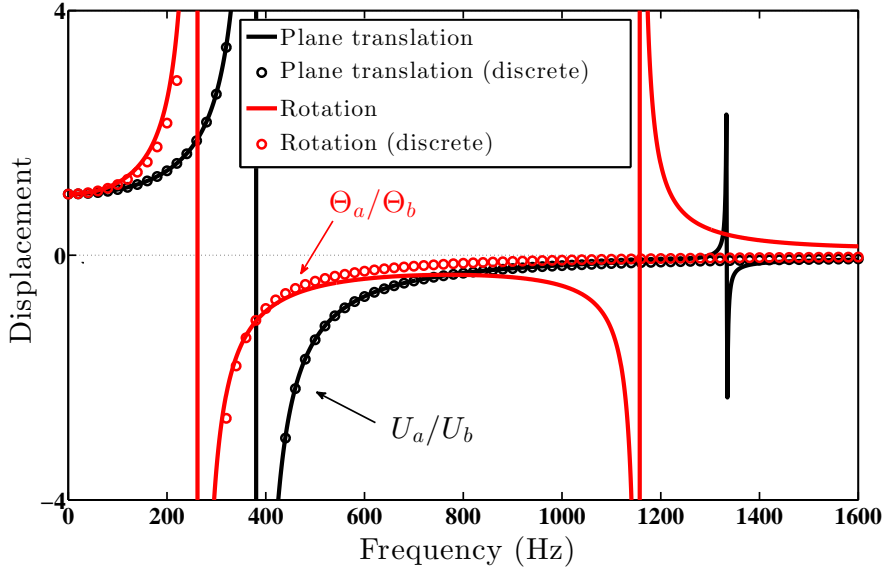} %
\caption{Displacement of the spherical core normalized to and exited by that of the matrix, based on the continuum model and discrete mass-spring model, and related to two modes: plane translation and rotation.}
\label{fig:H3D}
\end{figure}

Evolution of rotational displacement $H_*^{(r)}$, and  translational displacements of the lead sphere with frequency are shown in Fig. \ref{fig:H3D}, for a medium with the filling fraction 47\%, the lead sphere radius $a=  0.5$ cm, and the thickness of the  coating layer is $0.25$ cm. We note that, here also, with our material and geometrical  parameters, the first two local resonances of the rotational mode occur in lower frequencies compared to those of translational mode, as in the case of cylindrical inclusions.  The first resonance frequencies for the translation $f_0^{(p)}\approx 380$ Hz and  for rotational $ f_0^{(r)}\approx 261$ Hz. Similarly to the case of cylindrical system, the frequency of these resonances that correspond mainly to the vibration of the core sphere can be estimated through mass-spring models, leading to the following expressions   
\begin{eqnarray}
\label{eq:freq_3D}
\hspace*{-7mm}\lefteqn{\displaystyle f_0^{(p)}=f_0\left\lbrace 18 \frac{(1-\nu)(2-3\nu)\gamma^2}{(5-6\nu)(2-3\nu)(\frac{\gamma-1}{\gamma})-\frac{5}{4}\frac{(\gamma^2-1)^2}{\gamma^5-1}}\right\rbrace ^{1/2},
\;\displaystyle f_0^{(r)}=f_0\left( \frac{15\gamma^5}{\gamma^3-1}\right) ^{1/2}.} 
\end{eqnarray}
With our specific parameters, the above formulas give $f_0^{(p)}\approx 381$ Hz  and $ f_0^{(r)}\approx 272$ Hz, which are quite close to those obtained by the continuum model, in particular for the translation mode. Based on the mass-spring model, the displacement of the hard sphere can be obtained following the calculation of spring constants equivalent of the elastic layer [see Eqs. \eqref{eq:spring constants 3D} in Appendix \ref{appendix stiffness}]. Because of the physics of the second resonances in the cladding, we see here also from Fig. \ref{fig:H3D}, that the mass-spring model ignore these resonances. 
\begin{figure}
\hspace*{-10mm}\includegraphics[height=8 cm,draft=false]{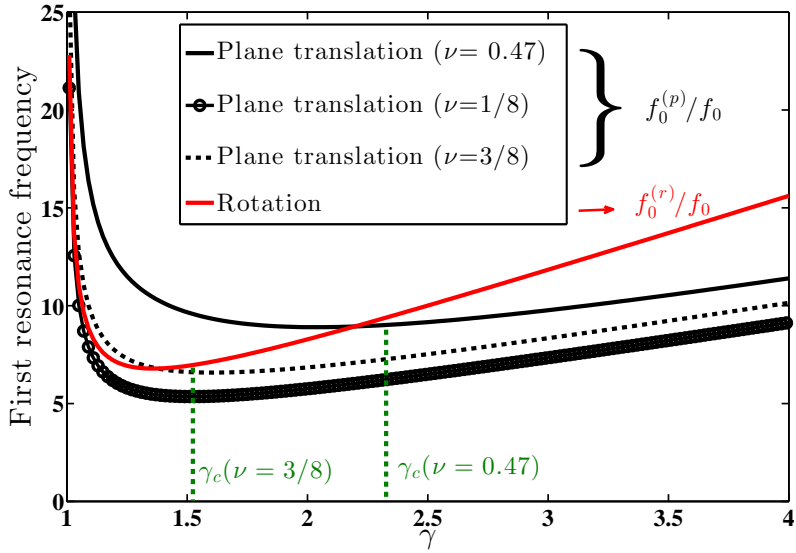} %
\caption{Evolution of the first resonance frequencies normalized by $f_0$ and estimated by the discrete mass-spring model [Eq. \eqref{eq:freq_3D}], with the form factor $\gamma=b/a$. The dependence of plane-translation mode on the Poisson's ratio $\nu$ of the coating  is also illustrated. The Poisson's ratio takes two representative values $1/8$ and $3/8$ in the intervals $[0, 1/4]$ and $[1/4, 1/2]$, respectively. The value of $\nu=0.47$ corresponds to the material parameters studied as an example throughout this paper (see Table \ref{table}). The rotational mode are independent of $\nu$.}
\label{fig:freq_3D}
\end{figure}

The study of the equations \eqref{eq:freq_3D} leads to the similar conclusions as in Sec. \ref{section cylinder}. Indeed, as long as $0\leq \nu\leq 1/4$, the rotational resonance frequency $f_0^{(r)}$ is higher than the translational resonance frequency $f_0^{(p)}$. For $1/4<\nu\leq 1/2$, the relative positions of these resonances  evolves with the parameter $\gamma=b/a$: for the small size inclusions $\gamma_c(\nu)<\gamma$, we have $f_0^{(p)}<f_0^{(r)}$, and vice versa otherwise (Fig. \ref{fig:freq_3D}). The quantity   $\gamma_c(\nu)$ is the unique solution to the equation $f_0^{(p)}(\gamma)=f_0^{(r)}(\gamma)$, which is an increasing function between $1$ for $\nu=1/4$, and   
$2.71$ when $\nu=1/2$. Fig. \ref{fig:freq_3D} shows the evolution of $f_0^{(p)}$ with $\gamma$ for three values of $\nu$: $1/8$ ($0 \leq \nu \leq 1/4$), $3/8$ ($1/4 <\nu \leq 1/2$), and $0.47$ that is the value corresponding to displacement behavior in Fig. \ref{fig:H3D}.

%

\section{Effective-medium properties: Metasolids}\label{section homog}

The knowledge of the kinematics of the cladding is of great interest to go forward in the dynamical homogenization of the whole structure, with the hard coated cylinder or sphere. Indeed, our purpose  is now to describe the homogenized behavior of the hard inclusion, the cladding and the matrix; and to obtain the key effective-parameters of the medium. 

\subsection{Homogenization for translational modes}

For a given translation of the embedding matrix $\bm{V}_b$, we aim at finding the equivalent density for the embedded solid (sphere or cylinder) and the cladding, subject to the induced translation $\bm{V}_a$. The knowledge of the stress distribution in the cladding permits us to evaluate the total force applied on the cladding by the matrix, and subsequently to derive the homogenized density. Indeed we can write 
\begin{equation}
-\rho_e \mathcal{V}_e\omega^2\bm{V}_b=\int\limits_{\Gamma_b}\bm{\sigma}.\bm{n}\,dS  \label{eq:Homo}
\end{equation}
where $\mathcal{V}_e$ is the volume of the region occupied by the hard inclusion $\Omega_a$ and the cladding $\Omega$, i.e. the region $\Omega_a\cup\Omega$, that is bounded by the surface $\Gamma_{b}$. The frequency-dependent effective  density $\rho_e$  is defined for each point inside $\Omega_a\cup\Omega$ and must coincide with the static effective value given by $\tilde{\phi}_a\rho_a+\tilde{\phi}\rho$, with $\tilde{\phi}_a$ and $\tilde{\phi}$ being the filling fraction of the coated solid and the cladding, respectively, 
such that $\tilde{\phi}_a+\tilde{\phi}=1$. 

In order to obtain the effective density for the whole unit cell, it remains to include the density of the matrix $\rho_b$, so that we can finally set $\rho_{\mathit{eff}}=(\phi+\phi_a)\rho_e+\phi_b\rho_b$, where $\phi_a$, $\phi$ and $\phi_b$ stand for the filling fraction of the coated solid, the cladding and the embedded matrix, respectively, such that  $\phi_a+\phi+\phi_b=1$. 

The case of spherical inclusions has  been studied in Ref. \cite{sheng2005}, through a result with an analytical expression \footnote{Note that there is a mistake in the given expressions of $g(\omega)$, $R(\omega)$, $g_1(\omega)$ and $g_2(\omega)$  [Eqs. (24), (25), (32) and (33) in this paper]. The terms $(T_{11}+T_{12})\mathrm{j}_1()+(T_{21}+T_{22})n_1()$ involved in $g(\omega)$ and $g_1(\omega)$ and $(T_{13}+T_{14})\mathrm{j}_1()+(T_{23}+T_{24})n_1()$ involved in $R(\omega)$ and $g_2(\omega)$ should be multiplied by $(1+(\alpha/\beta)^2)$. However, the graphics in Fig. 2 seem to be correct.}. Therefore, here we only deal with the cylindrical inclusions, where in particular an anisotropic aspect appears due to the fact that we take into account not only the plane-translation mode, but also the axial mode. Inserting $\bm{V}_b\equiv \bm{U}_b+\bm{T}_b$  in \eqref{eq:Homo}, the decoupling between the plane and axial translations leads  to two different expressions for the effective  density $\rho_e$, depending on whether we are dealing with the axial or plane translation. The anisotropy due to the geometrical configuration of the unit cell is obviously  the cause for the anisotropy of the effective mass density. We must  consider the effective density $\bm{\rho}_ {\mathit{eff}}$ as a second order tensor, and rewrite Eq. \eqref{eq:Homo} as $-\mathcal{V}_e \omega^2\bm{\rho}_{\mathit{eff}}. \bm{V_b}=\displaystyle\int_{\Gamma_b}\bm{\sigma}.\bm{n}\,dS$. Thus, the final expression will be 
\begin{eqnarray*}
\hspace*{-7mm}\lefteqn{\bm{\rho}_{\mathit{eff}}(\omega)=\rho_{\mathit{eff}}^{(p)}(\omega)\left(\bm{e}_x\otimes\bm{e}_x+\bm{e}_y\otimes\bm{e}_y\right) +\rho_{\mathit{eff}}^{(a)}(\omega)\bm{e}_z\otimes\bm{e}_z}
\end{eqnarray*}
with ($\beta=p,a$)
\begin{equation}
\rho_{\mathit{eff}}^{(\beta)}=(\phi+\phi_a)\rho_e^{(\beta)}+\phi_b\rho_b, 
\end{equation}
\begin{equation}
\rho_e^{(\beta)}(\omega)=\rho\left\lbrace g_1^{(\beta)}(\omega)+g_2^{(\beta)}(\omega)H^{(\beta)}(\omega)\right\rbrace ,
\end{equation}
\begin{subequations}
\begin{eqnarray*}
\hspace*{-7mm}\lefteqn{\begin{cases}
\;\displaystyle g_1^{(p)}(\omega)= 
\begin{pmatrix} [G(hb)] \\ [G(\kappa b)] \\ \end{pmatrix}^T
\begin{pmatrix} [E(ha)] & [F(\kappa a)] \\
		[E(hb)] & [F(\kappa b)] \\ \end{pmatrix}^{-1}
\begin{pmatrix} [0] \\ [1] \\ \end{pmatrix} \\
\\ 
\;\displaystyle g_2^{(p)}(\omega)=\frac{1}{\gamma}
\begin{pmatrix} [G(hb)] \\ [G(\kappa b)] \\ \end{pmatrix}^T
\begin{pmatrix} [E(ha)] & [F(\kappa a)] \\
		[E(hb)] & [F(\kappa b)] \\ \end{pmatrix}^{-1}
\begin{pmatrix} [1] \\ [0] \\ \end{pmatrix} \\
\end{cases}} \\
\hspace*{-7mm}\lefteqn{\begin{cases}
\;\displaystyle g_1^{(a)}(\omega)=\frac{2}{\kappa b}\frac{\mathrm{J}_0(\kappa
a)\mathrm{Y}_1(\kappa b)-\mathrm{J}_1(\kappa b)\mathrm{Y}_0(\kappa a)}{\mathrm{J}_0(\kappa a)\mathrm{Y}_0(\kappa
b)-\mathrm{J}_0(\kappa b)\mathrm{Y}_0(\kappa a)} \\ 
\\ 
\;\displaystyle g_2^{(a)}(\omega)=\frac{2}{\kappa b}\frac{\mathrm{J}_1(\kappa
b)\mathrm{Y}_0(\kappa b)-\mathrm{J}_0(\kappa b)\mathrm{Y}_1(\kappa b)}{\mathrm{J}_0(\kappa a)\mathrm{Y}_0(\kappa
b)-\mathrm{J}_0(\kappa b)\mathrm{Y}_0(\kappa a)} \\ 
\end{cases}}
\end{eqnarray*}
\end{subequations}
where the elements of the block matrices are defined in Appendix \ref{appendix expressions}. Fig. \ref{fig:Hom2D} shows the evolution of $\rho_{\mathit{eff}}^{(\beta)}(\omega)/\rho_{\mathit{eff}}^{(\beta)}(0)$  with  frequency, where we assume that $\rho_{\mathit{eff}}^{(\beta)}(0)$ matches with the static value $\rho_{0}\equiv\phi_a\rho_a+\phi\rho+\phi_b\rho_b$.
\begin{figure}
\hspace*{-5mm}\includegraphics[height=8 cm,draft=false]{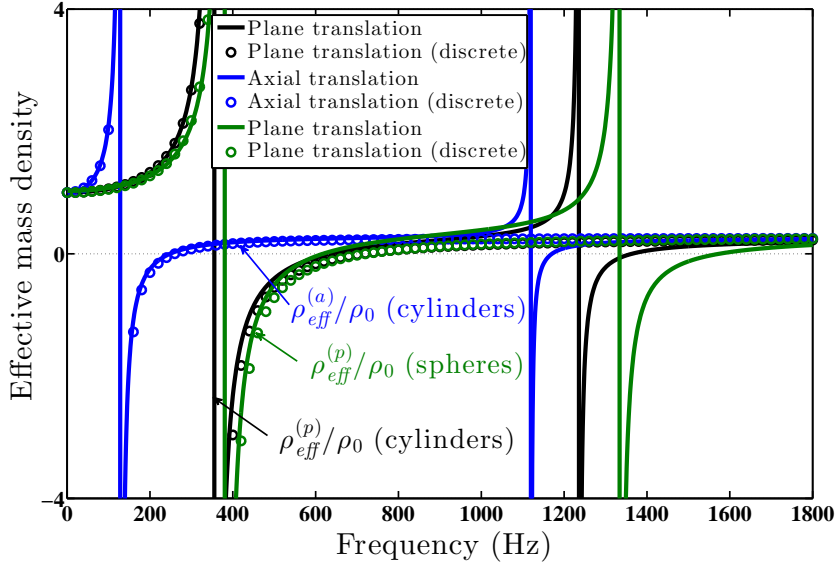} %
\caption{Effective mass densities based on continuum model and also discrete mass-spring model, associated with plane translation, axial translation, and rotation modes for media with cylindrical inclusions; and with plane translation and rotation modes for media with spherical inclusions.}
\label{fig:Hom2D}
\end{figure}
The evolutions for the axial or plane translations are qualitatively similar, but quantitatively different in that $\rho_{\mathit{eff}}^{(a)}$ becomes  negative at a lower frequency compared to $\rho_{\mathit{eff}}^{(p)}$ in accordance with the fact that the first resonance occurs for the axial mode (Fig. \ref{fig:Amp_2D}). We explained in Sec. \ref{section cylinder} that the (first) resonance of the axial should emerge at lower frequency, independently of the micro-structural parameters. This can be of particular interest, given that it is more challenging to make attenuate low-frequency waves. We note, however, that the resonance band where the mass density becomes negative, may also be less wide for the axial mode. With the present material and geometrical parameters, the band width for the axial mode  $\Delta f^{(a)}\approx 118$ Hz, while that of plane-translational mode $\Delta f^{(p)}\approx 280$ Hz.  

The effective masses related to the plane translations for media with cylindrical or spherical inclusions, and axial translation modes, can also be calculated through the discrete mass-spring model. Once the effective spring constants $k_{\mathit{eff}}^{(p)}$ (plane translation) and $k_{\mathit{eff}}^{(a)}$ (axial translation) are obtained (Appendix \ref{appendix stiffness}),  and the displacements $U_{a}/U_{b}$ (plane translation), $T_{a}/T_{b}$ (axial translation) calculated (Secs. \ref{section cylinder} and \ref{section sphere}), the equations of motion for the respective mass-spring systems lead easily to the frequency-dependent effective-masses for plane and axial translations. Fig. \ref{fig:Hom2D} shows these effective masses divided by the volume of the structural unit, in function of frequency. We see that  the discrete model predicts well the dynamic effective densities in low frequencies including the first negative region for both plane and axial translation modes. Expectedly, from the analysis on the resonance behaviors in Secs. \ref{section cylinder} and \ref{section sphere}, the second negative regions of effective mass densities that correspond to the second resonances (Figs. \ref{fig:Amp_2D} and \ref{fig:H3D}), are ignored by the discrete model for both modes and medium micro-geometries (Fig. \ref{fig:Hom2D}).


\subsection{Homogenization for rotational modes}

In order to obtain the effective parameter related to the rotations, we apply first the angular equation of motion to the sub-structure $\Gamma_{b\cup\Gamma}$ formed by the hard inclusion and the cladding. It reads

\begin{equation}
-I_e\omega^2\bm{\Theta}_b=\int\limits_{\Gamma_b}\bm{r}\times(\bm{\sigma}.\bm{n})\,dS
\label{eq:HomoR}
\end{equation}
where $I_e$ denotes the resulting moment of inertia around $\bm{e}_{z}$ for the embedded rigid body and its surrounded cladding that move in phase. Following a similar procedure regarding translation modes, we can  establish the expression of the density of the moment of inertia $i_e(\omega)$ such that $dI_e=i_ed\mathcal{V}_e$. Since this unusual quantity is obviously intensive in thermodynamic sense,  its homogenization in the medium should be possible. Tensorial behavior is involved to ensure the anisotropic aspects of the problem, hence we write : $-\omega^2\mathcal{V}_e  \bm{i}_e.\bm{\Theta}_b=\displaystyle\int_{\Gamma_b}\bm{r}\times(\bm{\sigma}.\bm{n})\,dS$. 

Similar to the case of translations, the formal expression of the effective density of the moment of inertia can be written as $\bm{i}_{\mathit{eff}}(\omega)=(\phi_a+\phi)\bm{i}_e(\omega)+\phi_bi_b\bm{\mathcal{I}}$. 

The spherical system gives rise to an effective density of moment of inertia $\bm{i}_{\mathit{eff}}\equiv i_{\mathit{eff}}\bm{\mathcal{I}}$, with 
\begin{equation}i_{\mathit{eff}}(\omega)=(\phi_a+\phi)i_e(\omega)+\phi_bi_b,
\end{equation} 
provided that 
\begin{equation}
i_{e}(\omega)=\frac{2}{5}\rho b^2\left\lbrace h^{*}_1(\omega)+h^{*}_2(\omega)H_*^{(r)}(\omega)\right\rbrace 
\end{equation} 
where, 
\begin{equation*}
\hspace*{-7mm}\begin{cases}
\;\displaystyle h^{*}_1(\omega)=\frac{5}{\kappa b}\frac{\mathrm{j}_1(\kappa a)\mathrm{y}_2(\kappa
b)-\mathrm{j}_2(\kappa b)\mathrm{y}_1(\kappa a)}{\mathrm{j}_1(\kappa a)\mathrm{y}_1(\kappa b)-\mathrm{j}_1(\kappa
b)\mathrm{y}_1(\kappa a)} \\ 
\\ 
\;\displaystyle h^{*}_2(\omega)=\frac{5}{\gamma\kappa b}\frac{%
\mathrm{j}_2(\kappa b)\mathrm{y}_1(\kappa b)-\mathrm{j}_1(\kappa b)\mathrm{y}_2(\kappa b)}{\mathrm{j}_1(\kappa
a)\mathrm{y}_1(\kappa b)-\mathrm{j}_1(\kappa b)\mathrm{y}_1(\kappa a)} \\ 
\end{cases}%
\end{equation*}
The isotropy of $\bm{i}_{\mathit{eff}}$ is clear and it is evident that $i_{\mathit{eff}}(\omega)$ coincides with the static value $i_{0}:=\phi_a i_a+\phi i+\phi_b i_b$ in the quasistatic limit $\omega\to 0$; with $\displaystyle i_a=\dfrac{2}{5}\rho_a a^2$, $\displaystyle i=\dfrac{2}{5}\rho \frac{b^5-a^5}{b^3-a^3}$, and $\displaystyle i_b=\dfrac{1}{6}\rho_b \frac{L^5-16\pi b^5/5}{L^3-4\pi b^3/3}$. 
\begin{figure}
\hspace*{-10mm}\includegraphics[height=9 cm,draft=false]{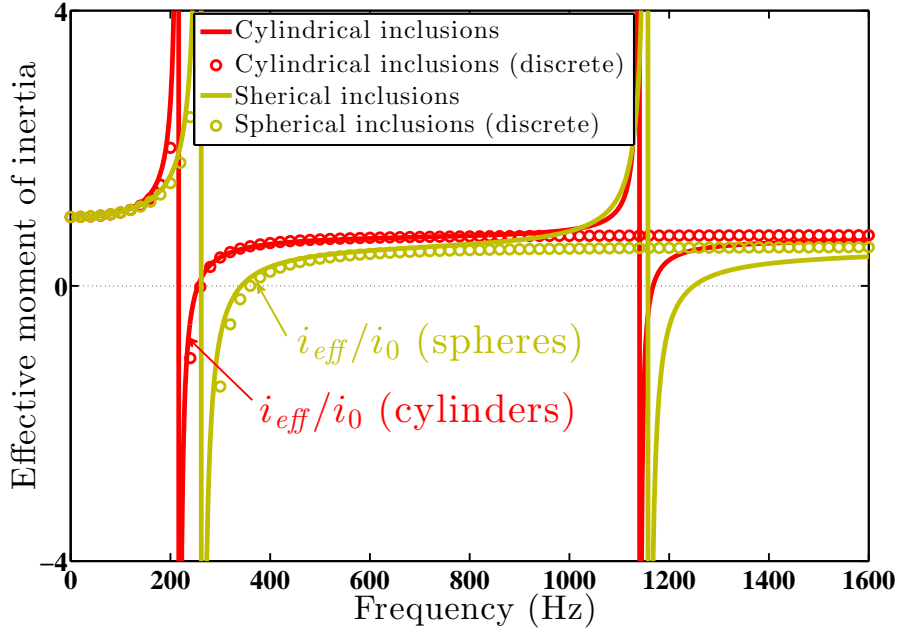} 
\caption{Density of moment of inertia based on continuum model and discrete mass-spring model, for media with cylindrical inclusions and spherical inclusions.}
\label{fig:moment inertia}
\end{figure}

Employing the same method that was used in the case of spherical system, for media with cylindrical inclusions, we obtain 
\begin{equation}
i_{e}(\omega)=\displaystyle\frac{1}{2}\rho b^2\left\lbrace h_1(\omega)+h_2(\omega)H^{(r)}(\omega)\right\rbrace 
\end{equation} 
with 
\begin{equation*}
\hspace*{-7mm}\begin{cases}
\;\displaystyle h_1(\omega)=\frac{4}{\kappa b}\frac{\mathrm{J}_1(\kappa a)\mathrm{Y}_2(\kappa
b)-\mathrm{J}_2(\kappa b)\mathrm{Y}_1(\kappa a)}{\mathrm{J}_1(\kappa a)\mathrm{Y}_1(\kappa b)-\mathrm{J}_1(\kappa
b)\mathrm{Y}_1(\kappa a)} \\ 
\\ 
\;\displaystyle h_2(\omega)=\frac{4}{\gamma\kappa b}\frac{
\mathrm{J}_2(\kappa b)\mathrm{Y}_1(\kappa b)-\mathrm{J}_1(\kappa b)\mathrm{Y}_2(\kappa b)}{\mathrm{J}_1(\kappa
a)\mathrm{Y}_1(\kappa b)-\mathrm{J}_1(\kappa b)\mathrm{Y}_1(\kappa a)} \\ 
\end{cases}%
\end{equation*}
where $\displaystyle i_a=\dfrac{1}{2}\rho_a a^2$, $\displaystyle i=\dfrac{1}{2}\rho (b^2+a^2)$, and $\displaystyle i_b=\dfrac{1}{6}\rho_b \frac{L^4-3\pi b^4}{L^2-\pi b^2}$. 

Fig. \ref{fig:moment inertia} presents the behavior of $i_{\mathit{eff}}$ in function of frequency, for media with cylindrical and spherical inclusions. This behavior with the negative values near rotational stop-band frequencies is similar to that we observed earlier for the effective mass densities. Here also, after calculating the spring constants for analogous mass-spring system (Appendix \ref{appendix stiffness}), and then obtaining the displacement $\Theta_{a}/\Theta_{b}$ (Figs. \ref{fig:Amp_2D} and \ref{fig:H3D}), the effective mass densities can be expressed according to the discrete model.  Again, from Fig. \ref{fig:moment inertia}  we observe that negative mass densities related to the lower resonance frequency band are appropriately described by the discrete model, while those associated with the higher frequency band, that is largely due to the rotational resonance inside the coating, are disregarded.  


\section{Summary and conclusions}\label{section conclusion}

A metasolid composed of a distribution of either cylindrical or spherical hard inclusions coated by a soft material has been studied. Generalizing an existing  continuum model \cite{sheng2005} that describes resonance-induced band gaps related to plain-translational modes, we provide analytical analysis to investigate anisotropic properties in the case of cylindrical inclusions, and also rotational modes for both cylindrical and spherical systems. 

We showed that the translational mode along the cylinders' axis and the rotational mode in both cylindrical and spherical systems could generate band gaps at low frequencies. The band gaps that originate in local resonance phenomena at long-wavelength regime, are tunable in terms of frequency by changing the coating's elastic properties and geometrical parameters of the structural unit. Furthermore, for all translational and rotational modes we found that the discrete mass-spring model could provide appropriate description of the material's kinematics including first resonances, but neglects the second resonances that occur at higher frequencies and mainly arise from resonances inside the coating.  The mass-spring model also provides simple analytical expressions allowing us to study easily the kinematics of first resonances in terms of the material parameters. 
We demonstrated that, while the macroscopic material dynamics is dictated by the  effective mass densities for translational modes, it is characterized  by its effective moment of inertia regarding the rotational modes.  Through analytical expressions, it was shown that components of frequency-dependent anisotropic effective densities for translational modes, and effective density of moment of inertia for rotational mode, become negative near corresponding resonance frequencies.  These effective parameters have been also calculated based on the mass-spring model, and their limits have been clarified by comparison to the continuum model.  

These results establish a step forward for metasolid characterization via introducing the density of moment of inertia as an effective-medium parameter. They give insight to improving the tunability of the metasolid for various applications, such as sound and vibration isolation,  and can serve as a physical-based quick guide for optimal design of the material to exhibit desired properties.

\section*{Acknowledgements}

The authors acknowledge support from the French National Research Agency (grant no. ANR-13-RMNP-0003-01 and ANR-11-LABX-022-01).

%

\appendix

\section{Expressions of the solutions to Helmholtz equations}
\label{appendix expressions}
Here, we will give the explicit form of the solutions to the Helmholtz equations \eqref{helmholtz equations}, combined with the boundary conditions for the case of cylindrical inclusions in Section \ref{section cylinder}. The boundary conditions \eqref{BC cylinder} and the solutions \eqref{solutions general form} lead to the following first conclusions
\begin{equation*}
\hspace*{-7mm}\begin{cases}
\;\ds C_n=0=D_n=G_n=H_n,\, \forall n\geq 2; \\
\;\ds A_0C_0=0=B_0C_0 \\
\;\ds E_0G_0\ \text{and}\ F_0G_0\ \text{to be determined}; \\
\;\ds A_1C_1,\,B_1C_1,\,E_1H_1,\ \text{and}\ F_1H_1\ \text{to be determined};\\ 
\;\ds A_1D_1,\,B_1D_1,\,E_1G_1,\ \text{and}\ F_1G_1\ \text{to be determined}.\\
\end{cases}
\end{equation*}

The expressions of the unknowns $E_0G_0$ and $F_0G_0$ are established in the article. The coefficients $A_1C_1$, $B_1C_1$, $E_1H_1$ and $F_1H_1$ involved in  \eqref{expression up cylinder} are solutions to the following linear system
\begin{equation}\label{linear system plane translation 1}
\underbrace{\begin{pmatrix}
[E(ha)] & [F(\kappa a)] \\
[E(hb)] & [F(\kappa b)] \\
\end{pmatrix}}_{\mathcal{D}}
\begin{pmatrix}
A_1C_1 \\
B_1C_1 \\
E_1H_1 \\
F_1H_1
\end{pmatrix}
= \begin{pmatrix}
[C_a] \\
[C_b] \\
\end{pmatrix},
\end{equation}
where the matrices used in the bloc-matrix formalism are 
\begin{eqnarray*}
\hspace*{-7mm}\lefteqn{[E(x)]=\begin{pmatrix}
x\mathrm{J}_1'(x) & x\mathrm{Y}_1'(x) \\
\mathrm{J}_1(x) & \mathrm{Y}_1(x) \\
\end{pmatrix}, \
[F(x)]=\begin{pmatrix}
\mathrm{J}_1(x) & \mathrm{Y}_1(x) \\
x\mathrm{J}_1'(x) & x\mathrm{Y}_1'(x) \\
\end{pmatrix}}\\
\hspace*{-7mm}\lefteqn{\text{and} \;
[C_x]=\begin{pmatrix}
xU_x\cos\theta_x \\
xU_x\cos\theta_x \\
\end{pmatrix}.}
\end{eqnarray*}

Similarly, the coefficients $A_1D_1$, $B_1D_1$, $E_1G_1$ and $F_1G_1$ in  \eqref{expression up cylinder} are determined by solving the linear system:
\begin{equation}\label{linear system plane translation 2}
\begin{pmatrix}
[E(ha)] & -[F(\kappa a)] \\
[E(hb)] & -[F(\kappa b)] \\
\end{pmatrix}
\begin{pmatrix}
A_1D_1 \\
B_1D_1 \\
E_1G_1 \\
F_1G_1
\end{pmatrix}
= \begin{pmatrix}
[S_a] \\
[S_b] \\
\end{pmatrix}
\end{equation}
with the same notations for the matrices $[E(x)]$ and $[F(x)]$, whereas $[S_x]=\begin{pmatrix}
xU_x\sin\theta_x \\
xU_x\sin\theta_x \\
\end{pmatrix}$.
The solutions of the above linear systems can be obtained with bloc-matrix formalism, which leads to  the final expression for the displacement field involved in \eqref{expression up cylinder}.

It is easy now to compute the ratio between the relative plane translations $\gr{U}_a$ and $\gr{U}_b$. Using \eqref{eq:PFD}, after some formal calculations, we obtain
\begin{equation}
\gr{U}_a= H^{(p)}(\omega)\; \gr{U}_b=\frac{-\gamma g^{(p)}(\omega)}{R^{(p)}(\omega)-\rho_a/\rho}\; \gr{U}_b
\label{eq:Hp}
\end{equation} 
with
\begin{equation}\label{gp and Rp}
\hspace*{-7mm}\begin{cases}
\;\ds g^{(p)}(\omega)= 
\begin{pmatrix} [G(ha)] \\ [G(\kappa a)] \\ \end{pmatrix}^T
\begin{pmatrix} [E(ha)] & [F(\kappa a)] \\
		[E(hb)] & [F(\kappa b)] \\ \end{pmatrix}^{-1}
\begin{pmatrix} [0] \\ [1] \\ \end{pmatrix} \\
\;\ds R^{(p)}(\omega)= 
\begin{pmatrix} [G(ha)] \\ [G(\kappa a)] \\ \end{pmatrix}^T
\begin{pmatrix} [E(ha)] & [F(\kappa a)] \\
		[E(hb)] & [F(\kappa b)] \\ \end{pmatrix}^{-1}
\begin{pmatrix} [1] \\ [0] \\ \end{pmatrix} \\
\end{cases}
\end{equation}
where $[G(x)]=\begin{pmatrix}\mathrm{J}_1(x) \\ \mathrm{Y}_1(x) \\ \end{pmatrix}$, $[0]=\begin{pmatrix}0 \\ 0 \\ \end{pmatrix}$, and $[1]=\begin{pmatrix}1 \\ 1 \\ \end{pmatrix}$.

We note that the relation \eqref{eq:Hp} is not only obtained for the plane translation in $\gr{e}_x$ direction (given after  solving \eqref{linear system plane translation 1}), that is  $U_a\cos\theta_a=H^{(p)}(\omega)U_b\cos\theta_b$, but also for the plane translation in $\gr{e}_y$ direction (given after solving \eqref{linear system plane translation 2}), i.e., $U_a\sin\theta_a=H^{(p)}(\omega)U_b\sin\theta_b$. Although this seems clear in the physical sense, the mathematical treatment is relatively tedious. 

%

\section{Expressions for effective stiffnesses according to mass-spring model}
\label{appendix stiffness}

The effective stiffnesses based on the discrete mass-spring model are obtained by using the static regime of general equations \eqref{Navier-w} for the cladding with $\omega=0$, and by imposing zero displacement on the inner boundary of the cladding $\Gamma_{a}$, with $\bm{V}_a=\bm{0}$ and $\bm{\Theta}_a=\bm{0}$, regarding Eq. \eqref{BC-w}. The resulting force gives the spring constants for translations (plane and axial), and the resulting torque leads to the spring constant for rotational motion. The corresponding equations are  
\begin{eqnarray*}
\hspace*{-6mm}\ds\int\limits_{\Gamma_a}\bm{\sigma}.\bm{n}\,dS=k_{\mathit{eff}}\bm{V}_b \;\; \text{and} \; \ds\int\limits_{\Gamma_a}\bm{r}\times (\bm{\sigma}.\bm{n})\,dS=C_{\mathit{eff}}\bm{\Theta}_b,
\end{eqnarray*}
for translational and rotational motions, respectively. In the above equations $k_{\mathit{eff}}$ is the spring constant, equivalent of the coating-layer that is represented by a spring in 1D tension/compression, and related to the (plane and axial) translational motion of the layer. Likewise, $C_{\mathit{eff}}$ is the spring constant describing the torsional stiffness of the coating layer that is taken to be analogous to a spring in 1D tension/compression, related to the rotational motion of the layer.  For cylindrical inclusions, the expressions of these constants are obtained as

\begin{eqnarray}
\label{eq:spring constants 2D}
\hspace*{-7mm}\lefteqn{\displaystyle k_{\mathit{eff}}^{(p)}=\mu\dfrac{8\pi L(1-\nu)(3-4\nu)}{(3-4\nu)^2\ln\gamma-\dfrac{\gamma^2-1}{\gamma^2+1}}, \; \displaystyle k_{\mathit{eff}}^{(a)}=\mu\dfrac{2\pi L} {\ln\gamma},
\;\displaystyle C_{\mathit{eff}}^{(r)}=\mu \dfrac{4\pi b^2 L}{\gamma^2-1},} 
\end{eqnarray}
where $ k_{\mathit{eff}}^{(p)}$, $k_{\mathit{eff}}^{(a)}$, and $C_{\mathit{eff}}^{(r)}$ refer to plane translation, axial translation, and rotation, respectively. For the spherical inclusions, we obtain 

\begin{eqnarray}
\label{eq:spring constants 3D}
\hspace*{-7mm}\lefteqn{\displaystyle k_{\mathit{eff}}^{(p)}=\mu\dfrac{24\pi a(1-\nu)(2-3\nu)}{(5-6\nu)(2-3\nu)(\dfrac{\gamma-1}{\gamma})-\dfrac{5}{4}\dfrac{(\gamma^2-1)^2}{\gamma^5-1}},
\;\displaystyle C_{\mathit{eff}}^{(r)}=\mu \dfrac{8\pi b^3}{\gamma^3-1}.}
\end{eqnarray}
The expressions for spring constants $k_{\mathit{eff}}^{(p)}$ and $C_{\mathit{eff}}^{(r)}$ for media with cylindrical and spherical inclusions can also be found in Ref. \cite{bonnet2015}.



\begin{thebibliography}{00}



\bibitem{sheng2000} Liu Z, Zhang X, Mao Y, Zhu Y Y,  Yang Z, Chan C T and Sheng P 2000 Locally resonant sonic Materials \textit{Science} \textbf{289} 1734

\bibitem{sheng2005} Liu Z, Chan C T  and Sheng P 2005 Analytic model of photonic crystals with local resonances \textit{Phys. Rev.} B \textbf{71} 014103

\bibitem{wu2008} Yang Z, Mei J, Yang M, Chan N H and Sheng P 2008 Membrane-type acoustic metamaterial with negative dynamic mass \textit{Phys. Rev. Lett.} \textbf{101} 204301


\bibitem{fang2006} Fang N, Xi D, Xu J, Ambati M, Srituravanich W, Sun C and Zhang X 2006 Ultrasonic metamaterials with negative modulus \textit{Nat. Mater.} \textbf{5} 452

\bibitem{nemati2015} Nemati N, Kumar A, Lafarge D and  Fang N X 2015 Nonlocal description of sound propagation through an array of Helmholtz resonators \textit{C. R. Mecanique} \textbf{343} 656


\bibitem{chan2004}  Li J and Chan C T 2004 Double-negative acoustic metamaterial \textit{Phys. Rev.} E \textbf{70} 055602(R)

\bibitem{ding2007} Ding Y, Liu Z, Qiu C and Shi J 2007 Metamaterial with simultaneously negative bulk modulus and mass density, \textit{Phys. Rev. Lett.} \textbf{99} 093904

\bibitem{cheng2008} Cheng Y, Xu J Y and Liu X J 2008 One-dimensional structured ultrasonic metamaterials with simultaneously negative dynamic density and modulus \textit{Phys. Rev.} B \textbf{77} 045134

\bibitem{lee2010} Lee S H, Park C M, Seo Y M,  Wang Z G and C. K. Kim 2010 Composite acoustic medium with simultaneously negative density and modulus \textit{Phys. Rev. Lett.} \textbf{104} 054301

\bibitem{liu2011} Liu X N, Hu G K,  Huang G L and  Sun C T 2011 An elastic metamaterial with simultaneously negative mass density and bulk modulus \textit{Appl. Phys. Lett.} \textbf{98} 251907 

\bibitem{sheng2013} Yang M, Ma G, Yang Z and Sheng P 2013 Coupled membranes with doubly negative mass density and bulk modulus \textit{Phys. Rev. Lett.} \textbf{110} 134301




\bibitem{wu2011} Wu Y, Lai Y and Zhang Z Q 2011 Elastic metamaterials with simultaneously negative effective shear modulus and mass density \textit{Phys. Rev. Lett.} \textbf{107} 105506 

\bibitem{sheng2011} Lai Y, Wu Y, Sheng P and Zhang Z Q 2011 Hybrid elastic solids \textit{Nat. Mater.} \textbf{10} 620 


\bibitem{brunet2015} Brunet T, Merlin A, Mascaro B, Zimny K, Leng J, Poncelet O, Aristégui C and  Mondain-Monval O 2015 Soft 3D acoustic metamaterial with negative index \textit{Nat. Mater.} \textbf{14} 384 


\bibitem{sigalas1993} Sigalas M and Economou E 1993 Band structure of elastic waves in two-dimensional systems \textit{Solid State Commun.} \textbf{86} 141 

\bibitem{djafari1993} Kushwaha M S, Halevi P, Dobrzynski L, Djafari-Rouhani B 1993 Acoustic band structure of periodic elastic composites \textit{Phys. Rev. Lett.} \textbf{71} 2022 


\bibitem{liang2012} Liang Z and Li J 2012 Extreme acoustic metamaterial by coiling up space  \textit{Phys. Rev. Lett.} \textbf{108} 114301 

\bibitem{cummer2013} Xie Y, Popa B I, Zigoneanu L and Cummer S A 2013 Measurement of a broadband negative index with space-coiling acoustic metamaterials \textit{Phys. Rev. Lett.} \textbf{110} 175501


\bibitem{milton2007} Milton G W and Willis J R 2007 On modifications of Newton's second law and linear continuum elastodynamics  \textit{Proc.  R. Soc.} A \textbf{463} 855 

\bibitem{cummer2007} Cummer S A and Schurig D 2007 One path to acoustic cloaking \textit{New J. Phys.} \textbf{9} 45 

\bibitem{chan2008} Ao X and Chan C T 2008 Far-field image magnification for acoustic waves using anisotropic acoustic metamaterials \textit{Phys. Rev.} E \textbf{77} 025601(R) 

\bibitem{zhang2009} Li J, Fok L, Yin X,  Bartal G and Zhang X 2009 Experimental demonstration of an acoustic magnifying hyperlens \textit{Nat. Mater.} \textbf{8} 931 

\bibitem{christensen2010} Zhu J, Christensen J, Jung J, Martin-Moreno L, Yin X,  Fok L, Zhang X and  Garcia-Vidal F G 2010 A holey-structured metamaterial for acoustic deep-subwavelength imaging \textit{Nat. Phys.} \textbf{7} 52  

\bibitem{zhu2011} Zhu X, Liang B, Kan W, Zou X and Cheng J 2011 Acoustic cloaking by a superlens with single-negative materials \textit{Phys. Rev. Lett.} \textbf{106} 014301
 
\bibitem{yu2014} Chen Y, Liu H, Reilly M, Bae H and  Yu M 2014  Enhanced acoustic sensing through wave compression and pressure amplification in anisotropic metamaterials \textit{Nat. Comm.} \textbf{5} 5247

\bibitem{djafari2015} Torrent D, Pennec Y and Djafari-Rouhani B 2015 Resonant and nonlocal properties of phononic metasolids \textit{Phys. Rev.} B \textbf{92} 174110 


\bibitem{zhao2005} Zhao H, Liu Y, Wang G, Wen J, Yu D, Han X and Wen X 2005 Resonance modes and gap formation in a two-dimensional solid phononic crystal \textit{Phys. Rev.} B \textbf{72} 012301

\bibitem{peng2012} Peng P, Mei J and Wu Y 2012 Lumped model for rotational modes in phononic crystals \textit{Phys. Rev.} B \textbf{86} 134304 

\bibitem{guenneau2013} Bigoni D, Guenneau S, Movchan A B and Brun M 2013 Elastic metamaterials with inertial locally resonant structures: Application to lensing and localization \textit{Phys. Rev.} B \textbf{87} 174303

\bibitem{huang2014} Zhu R, Liu X N, Hu G K, Sun C T and Huang G L 2014 Negative refraction of elastic waves at the deep-subwavelength scale in a single-phase metamaterial \textit{Nat. Commun.} \textbf{5} 5510 

\bibitem{peng2013} Peng P, Asiri S, Zhang X, Li Y and Wu Y 2013 A lumped model for rotational modes in periodic solid composites \textit{EPL} \textbf{104} 26001

\bibitem{bonnet2015} Bonnet G and Monchiet V 2015 Low frequency locally resonant metamaterials containing composite inclusions  \textit{J. Acoust. Soc. Am.} \textbf{137} 3263

\bibitem{Love44} Love A E H 1944 \textit{A Treatise on the Mathematical Theory of Elasticity} (New York: Dover)


 

\bibitem {bonnet2017} Bonnet G and Monchiet V 2017 Dynamic mass density of resonant metamaterials with homogeneous inclusions \textit{J. Acoust. Soc. Am.} \textbf{142} 890
 
\bibitem {huang2009} Huang H H and Sun C T Wave attenuation mechanism in an acoustic metamaterial with negative effective mass density 2009 \textit{New J. Phy.} \textbf{11} 013003 
   


\end{thebibliography}
\end{document}